\documentclass[10pt]{iopart}
\usepackage{graphicx}
\usepackage{epstopdf} 
 \expandafter\let\csname equation*\endcsname\relax
  \expandafter\let\csname endequation*\endcsname\relax
\usepackage{amsmath}
\usepackage{amsfonts}
\usepackage{iopams}
\usepackage{mathtools}
\usepackage{braket}
\usepackage{ulem}
\usepackage{color}
\usepackage{amstext}

\begin{document}

\title{Catastrophes in non-equilibrium many-particle wave functions: universality and critical scaling}

\author{J Mumford, W Kirkby and D H J O'Dell}

\address{Department of Physics and Astronomy, McMaster University, 1280 Main St.\ W., Hamilton, ON, L8S 4M1, Canada}

\begin{abstract}
As part of the quest to uncover universal features of quantum
dynamics,  we study catastrophes that form in simple many-particle
wave functions following a quench, focusing on two-mode systems that
include the two-site Bose Hubbard model, and under some circumstances
optomechanical systems and the Dicke model. When the wave function is
plotted in Fock space certain characteristic shapes, that we identify as cusp catastrophes,
appear under generic conditions. In the vicinity
of a cusp the wave function takes on a universal structure
described by the Pearcey function and obeys scaling relations which
depend on the total number of particles $N$. In the thermodynamic
limit ($N \rightarrow \infty$) the cusp becomes singular, but at
finite $N$ it is decorated by an interference pattern. This pattern
contains an intricate network of vortex-antivortex pairs, initiating a
theory of topological structures in Fock space. In the case where the
quench is a $\delta$-kick the problem can be solved analytically and we obtain
scaling exponents for the size and position of the cusp, as well as
those for the amplitude and characteristic length scales of its
interference pattern.  Finally, we use these scalings to describe the
wave function in the critical regime
of a $\mathbb{Z}_2$ symmetry-breaking dynamical phase transition.
\end{abstract}



\date{\today}

\maketitle

\ioptwocol

\section{Introduction}
Universality is one of the most cherished concepts in physics. Perhaps the best known example is  near second-order (continuous) phase transitions where \textit{equilibrium} properties such as correlation lengths and susceptibilities diverge according to power laws with universal exponents as a control parameter approaches its critical value. In fact, physical systems are partitioned into different universality classes, each characterized by a particular set of critical exponents. Members of the same class can be very different at the microscopic scale and yet they display the same asymptotic scale invariance in the critical regime. 

Our goal in this paper is to study universality in non-equilibrium behaviour.  Current paradigms in this area include the Kibble-Zurek mechanism \cite{kibble76,zurek85} describing defect production upon ramping through a second order phase transition at a finite speed, and the eigenstate thermalization hypothesis describing thermalization of isolated quantum systems \cite{deutsch91,srednicki94,rigol08}. These problems have attracted the attention of the cold atom \cite{greiner02,lamacraft07,kollath07,rossini09,trotzky12,gerving12,torre13,braun15,nicklas15,navon15,eisert15,anquez16} and cold ion \cite{campo10,ejtemaee13} communities because such systems offer remarkable levels of coherence and control, making them useful for testing fundamental models of many-particle dynamics.  

The universality we investigate here is somewhat different and occurs in the time-dependent many-particle wave function itself (rather than, say, correlation functions). In particular, we study striking geometric shapes that emerge in Fock space following a quench, identifying them as the catastrophes that are categorized by catastrophe theory (CT) \cite{thom75,arnold75,poston12}. They can occur far from any phase transition, although close to one they display familiar features such as critical slowing down.   Catastrophes do in fact have a number of features that are reminiscent of phase transitions, including the occurrence of singularities, equivalence classes, and self-similar scaling relations \cite{berry77,varchenko76}.

 A list of the structurally stable catastrophes with co-dimension one, two and three is given in Table \ref{tab:catastrophes}.  Each is defined via its normal form or generating function $\Phi(\mathbf{s};\mathbf{R})$; each generating function is a polynomial in the state variables $\mathbf{s}=\{s_{1},s_{2},s_{3},\ldots \}$ but is linear in the control parameters $\mathbf{R}=\{X,Y,Z,\ldots \}$. \textit{In this paper the physical role of the generating function is as the mechanical action}. In this way, each canonical generating function is associated with a canonical wave function via a Feynman path integral $\Psi(\mathbf{R}) \propto \int \exp [\rmi \Phi(\mathbf{s}; \mathbf{R})/\hbar ] \, \rmd \mathbf{s}$   \cite{duistermaat74,berry76}. The state variables $\mathbf{s}$ specify the ``paths'' or configurations and the control parameters $\mathbf{R}$ provide the coordinates.   In the simplest case of the fold catastrophe this gives the Airy function \cite{airy1838}, and in the case of the cusp, which will be the main subject of this paper, it gives the Pearcey function \cite{pearcey46}. These functions, referred to variously as wave catastrophes or diffraction integrals \cite{berry81,nye99}, have the status of special functions akin to, say, Bessel functions, and their mathematical properties are summarized in chapter 36 of reference \cite{NISThandbook}.  In a typical physical problem the action does not automatically present itself in one of the normal forms listed in Table \ref{tab:catastrophes}, but the claim of CT is that close to a singularity it can always be mapped onto one of them. Finding the required transformation may not be easy, but in the present paper we shall consider simple situations where this can be done analytically.

 It is important to point out that catastrophe theory can be applied in a number of different ways to quantum mechanics. Our use of the catastrophe generating functions $\Phi(\mathbf{s};\mathbf{R})$ as actions is distinct from other applications, such as taking the generating functions as potentials to be used in the Schr\"{o}dinger equation \cite{Gilmore86,Emary05}, although in both cases universal structures are obtained which have a qualitative robustness. This important property, which is known as structural stability, means that catastrophes are qualitatively immune to perturbations and hence occur generically with no need for special symmetry. This is the reason behind their ubiquity.

 \begin{table*}
\caption{\label{tab:catastrophes} Structurally stable catastrophes and their generating functions with co-dimension $K \le 3$. Co-dimension is defined as the dimensionality of the control space minus the dimensionality of singularity. $\mathbf{R}$ represents the control parameters and $\mathbf{s}$ the state variables.}
\begin{indented}
\item[]\begin{tabular}{@{}lll}
\br
name & $K$ & generating function $\Phi(\mathbf{s}; \mathbf{R})$  \\
\mr
fold &1 & $s^3 + X s$ \\
cusp & 2 & $s^4+X s^2+Y s$ \\
swallowtail & 3 & $s^5+X s^3+Y s^2+Z s$ \\
elliptic umbilic & 3 & $s_{1}^3 - 3 s_{1} s_{2}^2 + Z (s_{1}^2+s_{2}^2) + Y s_{2} + X s_1$ \\
hyperbolic umbilic & 3 & $s_{1}^3+s_{2}^3 + Z s_{1} s_{2} +Y s_{2} + Z s_{1}$ \\
\br
\end{tabular}
\end{indented}
\end{table*}

Our application of catastrophe theory  in this paper is inspired by its use in the description of optical caustics \cite{berry81,nye99,berry80}. 
Caustics are the result of natural focusing  and occur widely in nature with examples including rainbows, bright lines on the bottom of swimming pools, twinkling of starlight \cite{berry77}, gravitational lensing, and freak waves \cite{hohmann10}.  Being a general wave phenomenon, caustics also appear in quantum waves such as those describing the motion of cold atoms. The experiment by Rooijakkers {\it et al} \cite{rooijakkers03} observed caustics in the trajectories of cold atoms trapped in a magnetic waveguide, Huckans {\it et al} \cite{huckans09} observed them in the dynamics of a Bose-Einstein condensate (BEC) in an optical lattice, and in the experiment by Rosenblum  {\it et al} \cite{rosenblum14} caustics appeared when a cold atomic cloud was reflected from an optical barrier in the presence of gravity. On the theoretical side, caustics have been predicted to occur in atomic diffraction from standing waves of light \cite{odell01}, in atom clouds in pulsed optical lattices \cite{averbukh01,leibscher04}, in the dynamics of particles with long-range interactions \cite{barre02}, in the expansion dynamics of Bose gases released from one- and two-dimensional traps \cite{chalker09}, and they can also produce characteristic features in the long-time (but non-thermal) probability distribution following quenches in optical lattices and Josephson junctions \cite{berry99,odell12}. Furthermore, although not identified as such by their authors, caustics can be seen in figures in papers on the dynamics of BECs encountering a supersonic obstacle \cite{carusotto06}, on the collapse and subsequent spreading of a BEC of polaritons pulsed by a laser \cite{dominici15} and in quantum random walks by interacting bosons in an optical lattice \cite{preiss15}.

The properties of caustics depend on the scale at which they are viewed. At large scales they appear singular and the proper description is via geometric ray theory, equivalent to the classical ($\hbar \rightarrow 0$) limit of single-particle quantum mechanics. In this theory the intensity tends to infinity as the caustic is approached. At small scales, where the wavelength is finite, the singularity is removed by interference. Each class of caustic is dressed by a characteristic  interference pattern (wave catastrophe). In the many-particle problem there are two new features: the first is a rather trivial replacement of  $\hbar$ by $1/N$, where $N$ is the total number of particles. The second, more fundamental difference, is an intrinsic granularity imposed on wave catastrophes by the discreteness of the number of particles \cite{odell12}. This latter feature is particularly apparant in Fock space which is the natural arena for many-particle physics. In many-particle problems mean-field theory plays the role of geometric ray theory: it applies in the limit $N \rightarrow \infty$ and ignores the granularity of the particle number, providing an effective single-particle description which is usually nonlinear.  

As an example, consider a BEC containing $N$ ultracold atoms.  In the mean-field theory for condensed bosons the condensate wave function $\psi(\mathbf{r},t)$ obeys the Gross-Pitaevskii wave equation (GPE)
\begin{equation}
i \hbar \frac{\partial \psi}{\partial t} = \left ( -\frac{\hbar^2}{2
    m} \nabla^2 +V(\mathbf{r}) + g \vert \psi \vert^2 \right ) \psi \ ,
\label{eq:GPE}
\end{equation}
where $V(\mathbf{r})$ is the external potential and $g$ characterizes the strength of the interactions. This `first-quantization' in terms of a classical wave equation is sufficient to remove singularities in coordinate space. However, in Fock space mean-field theory predicts singular caustics that must be removed by second-quantizing the field, i.e.\ by building in the discreteness of the number of field quanta (atoms) which is ignored by the GPE  (see Figure \ref{fig:BJJ} below) \cite{odell12}. In this paper we shall work in the semiclassical regime ($N \gg 1$) where a continuum approximation can be applied to Fock space although crucially we retain the non-commuting nature of quantum operators (such as the number and phase operators), in contrast to the mean-field approximation. Under this prescription standard continuous wave catastrophes are recovered \cite{mumford16}.

A singularity in Fock space can be considered to be an example of a {\it quantum catastrophe}, i.e.\ a singularity in classical field theory that is removed by going over to quantum field theory where the field amplitudes are quantized (atoms in the case of BECs, photons in the case of electromagnetic fields \cite{berry04,berry08}). Hawking radiation, where pairs of photons are produced from the vacuum near a black hole, is an example of a quantum catastrophe as has been pointed out by Leonhardt \cite{leonhardt02} by considering the fate of a classical electromagnetic wave propagating over an event horizon. The wave suffers a phase singularity (it oscillates infinitely rapidly and hence takes all values) when seen by an observer at infinity. Indeed, there is no Hermitian operator for phase in quantum mechanics and the concept of phase only becomes well defined in the classical limit of a large number of quanta.

In this paper we study simple many-particle systems involving just two modes. This includes the two-site Bose-Hubbard model (a particular case of the Lipkin-Meshkov-Glick model \cite{lipkin65}), the Dicke model, various optomechanical systems, and generally any collection of spins or pseudo-spins in the single mode approximation (including the Ising model with long-range interactions \cite{das06}).  As we shall show in Section \ref{sec:approx}, in the semiclassical regime  these models can be mapped onto an effective Hamiltonian of the form
\begin{equation}
\frac{\hat{H}}{N} = \frac{\hat{p}^2}{2} + V(\hat{x}),
\label{eq:ham}
\end{equation}
where $V(\hat{x})$ is an operator with a non-linear (anharmonic) spectrum.  Since this Hamiltonian has one degree of freedom, the space where dynamical catastrophes live is the two-dimensional $(x,t)$-plane known as the control space, and according to CT the structurally stable catastrophes in two dimensions are fold lines which can meet at cusp points (a general feature of CT is that the higher catastrophes contain the lower ones). We therefore expect from the very start that the structures we see will be comprised of Airy and Pearcey functions.  Furthermore, all these models display second-order phase transitions as a parameter is varied and this fact will allow us to examine how catastrophes behave when the Hamiltonian is tuned close to the critical point.

The plan for the rest of this paper is as follows: After reviewing some examples of two-mode many-particle systems in Section \ref{sec:approx}, we proceed in Section \ref{sec:cat} to study the classical (mean-field) dynamics of these systems following a quench, showing how catastrophes arise as the envelopes of families of classical trajectories compatible with the quantum conditions. Specializing to the $\delta$-kicked case in Section \ref{sec:kickedHamiltonians}, we demonstrate the connection between the second-order phase transition in the instantaneous Hamiltonian and the appearance in the subsequent dynamics of different types of cusp catastrophe in Fock space + time. In Section \ref{sec:quantum} we examine the quantum version of this behaviour, showing how the wave function can be mapped onto the Pearcey function. This function obeys a set of scaling identities and we use these to understand the scaling properties of the many-particle wave function, including the size and position of the cusp, the oscillations in the interference pattern that decorates it, as well as topological features such as vortices. In Section \ref{sec:limitations} we look beyond the $\delta$-kicked case and discuss the features we expect when the system propagates under the full Hamiltonian. We give our conclusions in Section \ref{sec:conclusion}.

The results presented in Sections \ref{sec:approx} and \ref{sec:cat} are largely review, with the idea that granular catastrophes appear in the Fock space of many-particle systems being introduced previously by one of us (DO) in \cite{odell12}. However, the mapping presented in Sections \ref{sec:kickedHamiltonians},  \ref{sec:quantum} and  \ref{sec:limitations} of $\delta$-kicked two-mode many particle wave functions onto the Pearcey function is to the best of our knowledge new, including the connection to dynamical phase transitions and the concept of quantized vortices in Fock space.

\section{Two-Mode Many-Particle Systems}
\label{sec:approx}

In this section we show how various two-mode many-particle Hamiltonians can be written in the form given in Eq.\ (\ref{eq:ham}).  The Hilbert space of Eq.\ (\ref{eq:ham}) is infinite, so it cannot  properly model highly excited states that feel the finiteness of the original Hilbert space, however, when $N$ is large and the highest states are not excited Eq.\ (\ref{eq:ham}) can be used as a
semiclassical approximation.  Because $1/N$ plays the role of $\hbar$,  the operators $\hat{x}$ and $\hat{p}$ satisfy the commutation relation $[\hat{x},\hat{p}] = \mathrm{i}/N$, and the
classical limit $\hbar \rightarrow 0$ is the same as the thermodynamic limit $N \rightarrow \infty$. Away from this limit, the finite value of the commutator must be preserved if singular caustics are to be avoided in Fock space.

\subsection{Two-site Bose-Hubbard model}

We begin with the Bose-Hubbard model with two sites and $N$ particles. This can be used to describe a BEC in a double well potential which has been realized in a number of experiments \cite{albiez05,schumm05,levy07,zibold10,leblanc11,trenkwalder16}.
In the single band regime the two modes can be taken to be the ground states on each site and the Hamiltonian is written \cite{gati07}
\begin{equation}
\hat{H}_{\mathrm{BH}} = U \hat{n}^2 - J \left ( \hat{a}_R^\dagger
\hat{a}_L+ \hat{a}_L^\dagger \hat{a}_R \right )
\label{eq:BH}
\end{equation}
where $\hat{n} = (\hat{a}^\dagger_R \hat{a}_R -
  \hat{a}^\dagger_L \hat{a}_L )/2$ is half the
number-difference between the two sites labeled by $L$ (left) and  $R$ (right).  The
annihilation and creation operators obey the usual bosonic commutation
relations $[\hat{a}_{L/R},\hat{a}^{\dag}_{L/R}]= \delta_{L/R}$.  $U$ is the on-site interaction
energy between the bosons and can be positive or negative depending upon whether the interactions are repulsive or attractive, and $J > 0$ is the intersite hopping energy. The parameter $\Lambda_{\mathrm{BH}} = UN/2J$,  which is the ratio of the interaction energy to the mode-coupling energy, determines the behaviour of the system. For attractive enough interactions,  $\Lambda_{\mathrm{BH}} < -1$, the ground state suffers a $\mathbb{Z}_2$ symmetry-breaking phase transition where a majority of bosons clump on one site or the other, as seen in the recent experiment by Trenkwalder {\it et al} \cite{trenkwalder16}. When $\Lambda_{\mathrm{BH}} > -1$ the ground state is symmetric but the dynamics can be divided into three regimes \cite{gati07}: the Rabi regime ($-1 < \Lambda_{\mathrm{BH}} < 1$) where the interactions (which provide the nonlinearity) are weak enough that the system essentially behaves as $N$ independent two-level oscillators (pseudo-spins); the Josephson regime ($1 <\Lambda_{\mathrm{BH}} \ll N^2$)
where both the interactions and the single particle hopping are important; and the Fock regime ($\Lambda_{\mathrm{BH}} \gg N^2$) where interactions dominate. 

The many-particle 
wave function can be expanded $\vert \Phi (t) \rangle=\sum c_{n}(t) \vert n \rangle$ in terms of the eigenstates $\vert n \rangle$ of $\hat{n}$, i.e.\ in terms of Fock states with well defined number differences. In general the system is in superposition of number difference states and the probabilities $\vert c_{n}(t) \vert^2= \vert \langle n \vert \Phi (t) \rangle \vert^2$ define the probability distribution in Fock space. There is no explicit assumption of BEC although the bosons must be cold enough to only occupy the lowest state on each site. By contrast, in the Gross-Pitaevskii mean-field theory it is assumed that there is condensate on each site with a perfectly well-defined number difference $n(t)$ and phase difference $\phi(t)=\phi_{R}(t)-\phi_{L}(t)$ between the two sites at all times \cite{pitaevskii01}, in other words $\Delta n \Delta \phi =0$. This implies a U(1) symmetry breaking in which the phase difference is selected. Furthermore, the number difference becomes a continuous variable rather than a discrete one. The mean-field Hamiltonian is \cite{smerzi97}
\begin{equation}
\lim_{N \rightarrow \infty} \frac{\hat{H}_{\mathrm{BH}}}{NJ} = H_{\mathrm{BH}} = \Lambda_{\mathrm{BH}} \frac{z^2}{2} - \sqrt{1-z^2}\cos\phi \, .  \label{eq:BJJHam}
\end{equation}
where it is customary to introduce $z = 2n/N$, where $-1 \le z \le 1$, as the number difference scaled by the total number of bosons. This Hamiltonian corresponds to that of a pendulum where the role of the angular momentum is played by the number difference and its angular position is given by the phase difference. However, the length of the pendulum depends on its angular momentum via the square root factor which gives rise to a type of classical motion, called $\pi$-oscillations, that is not present in the rigid pendulum  \cite{smerzi97}. In the Rabi regime there are two stable stationary points, one at $\phi=0$ and the other at $\phi=\pi$, the latter corresponding to the pendulum standing upright.  Small oscillations around $\phi=0$ are called plasma oscillations (in analogy to similar excitations in Josephson junctions) and were observed using cold atoms in the pioneering experiments by Albiez {\it et al} \cite{albiez05} and Levy {\it et al} \cite{levy07}. $\pi$-oscillations, on the other hand, correspond to small oscillations around $\phi=\pi$ and were seen in the experiment by Zibold {\it et al} \cite{zibold10}. Both plasma and $\pi$-oscillations have a time-averaged number difference of $\langle z \rangle = 0$ but are distinguished by having a time-averaged phase differences of 
$\langle \phi \rangle = 0$ and $\langle \phi \rangle = \pi$, respectively. However, upon entering the Josephson regime there is pitchfork bifurcation in which the stationary point at $\phi=\pi$ becomes unstable and is replaced by two new stable stationary points which have $\langle z \rangle \neq 0$. These excited yet stationary states are responsible for the phenomenon of macroscopic quantum self-trapping \cite{raghavan99}  where an initial imbalance of boson number between the two wells remains locked in place (rather than oscillating back and forth) and is related to the Josephson ac effect in Josephson junctions. The stationary point at $\phi=0$ is unaffected by the bifurcation but is separated from the new stationary points by a separatrix. In the quantum theory the separatrix corresponds to a peak in the density of states \cite{krahn09} and can be interpreted as a dynamical phase transition in the thermodynamic limit \cite{santos16}. The transition is of the $\mathbb{Z}_2$ symmetry breaking type corresponding to the choice of either $\langle z \rangle > 0$ or $\langle z \rangle < 0$.

 The quantum dynamics is governed by the Schr\"{o}dinger equation
$i \hbar \partial_{t} \vert \Phi \rangle =\hat{H}_{\mathrm{BH}} \vert \Phi \rangle$. Substituting in the expansion over Fock states one obtains a set of $N+1$ coupled differential-difference equations  for the Fock space amplitudes $c_n(t)$. These can easily be solved numerically \cite{krahn09}, and can also be tackled analytically in the semiclassical regime \cite{odell01,odell12} revealing cusp catastrophes in the wave function in Fock space plus time following a quench. The cusps have also been discussed in terms of quantum collapses and revivals of the initial state \cite{milburn97,veksler15}.

\begin{figure}[ht]
\begin{center}
\includegraphics[width=1\columnwidth]{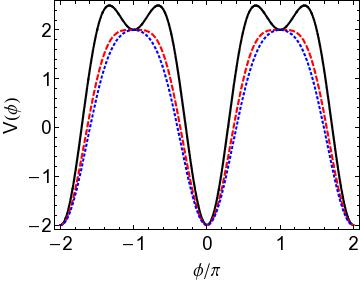}
\end{center}
	\caption{A plot of the effective potential $V(\phi)$ for the two-site Bose-Hubbard model as given in Eq.\ (\ref{eq:V(phi)}). Each curve is for a different value of  $\Lambda_{\mathrm{BH}}$: 0.5 (solid black), 1.0 (dashed red), and 1.5 (dotted blue). When $\Lambda_{\mathrm{BH}}<1$ there are two minima, one at $\phi=0$ and the other at $\phi=\pm \pi$, and motion about these points gives rise to plasma- and $\pi$-oscillations, respectively. When $\Lambda_{\mathrm{BH}}>1$ the potential features only a single minimum at $\phi=0$. Two types of motion are possible in this latter regime: plasma oscillations around the minimum and macroscopic quantum self-trapping where the energy exceeds the barrier top at $\phi =\pm \pi$ and the phase grows continuously, either in the clockwise or anticlockwise direction. Note that $V(\phi)$ is periodic outside the fundamental domain $- \pi < \phi \le \pi $, but we have plotted twice this range so that the properties of the potential near $\phi = \pm \pi$ are clear. } 
	\label{fig:phipot}
\end{figure}

For the purposes of this paper we seek a semiclassical Hamiltonian in the form of Eq.\ (\ref{eq:ham}). The mean-field Hamiltonian given in Eq.\ (\ref{eq:BJJHam}) is close to the desired structure and can be re-quantized by promoting $z$ and $\phi$ to operators. However, in contrast to the original problem, we now assume that $\hat{z}$ (and $\hat{\phi}$) has a continuous spectrum and obeys the commutation relation $[\hat{\phi},\hat{z}] = 2 \mathrm{i}/N$ \cite{krahn09}. We refer to this as the \textit{continuum approximation}. There is still the matter of the square root factor involving $\hat{z}$ which means that this Hamiltonian is not quite separated into ``kinetic plus potential energy''. To remedy this we write the wave function (in the phase representation) as  
\begin{equation}
\Psi(\phi) =     e^{ \frac{N}{2 \Lambda_{\mathrm{BH}}}  \cos \phi}  \sum_{n = -
  N/2}^{N/2} \frac{c_n}{\sqrt{\left ( \frac{N}{2} + n \right )! \left
      ( \frac{N}{2} - n \right )!}} e^{i n \phi} .
\end{equation}
Note that this wave function is not normalized.
The time-independent Schr\"{o}dinger equation then becomes (in the semiclassical regime $N \gg
1$) \cite{ulyanov92}
\begin{equation}
- \frac{2 \Lambda_{\mathrm{BH}}}{N} \frac{\partial^2 \Psi}{\partial \phi^2} -
\frac{N}{2 \Lambda_{\mathrm{BH}}} \left ( \mathrm{cos}^2 \phi + 2 \Lambda_{\mathrm{BH}}
  \mathrm{cos} \phi -1 \right ) \Psi = \frac{E}{J} \Psi \, ,
\end{equation}
where $ \hat{z} = -\frac{2i}{N} \frac{\partial}{\partial \phi}$ in analogy to the standard relation $\hat{p}=-i \hbar \partial/ \partial x$. This suggests the effective Hamiltonian
\begin{equation}
\frac{\hat{H}^\prime_{BH}}{NJ} = \frac{\Lambda_{\mathrm{BH}}}{2} \hat{z}^2 -
\frac{1}{2 \Lambda_{\mathrm{BH}}} \left ( \mathrm{cos}^2 \hat{\phi} + 2 \Lambda_{\mathrm{BH}}
  \mathrm{cos} \hat{\phi} -1 \right )
\label{eq:approxBH}
\end{equation}
where we use the prime to signify that Eq.\ (\ref{eq:approxBH}) is the re-quantized
 version of Eq.\ (\ref{eq:BH}). Equation (\ref{eq:approxBH}) has the form of Eq.\ (\ref{eq:ham}) where 
 \begin{equation}
 V(\hat{\phi})=-\frac{1}{2 \Lambda_{\mathrm{BH}}} \left( \mathrm{cos}^2 \hat{\phi} + 2 \Lambda_{\mathrm{BH}}
  \mathrm{cos} \hat{\phi} -1 \right)
  \label{eq:V(phi)}
 \end{equation} 
 plays the role of an effective potential for the position coordinate $\phi$ \cite{raghavan99} which we plot in Figure \ref{fig:phipot}. When $\Lambda_{\mathrm{BH}}<1$ we see two minima, one at $\phi=0$ and the other at $\phi= \pm \pi$, which are responsible for the plasma and $\pi$-oscillations, respectively. As expected, the minimum at $\phi= \pi$ disappears at $\Lambda_{\mathrm{BH}}=1$ corresponding to the destruction of the $\pi$-oscillations. When $\Lambda_{\mathrm{BH}}>1$ the potential has just a single well and two types of motion are possible: when the energy is below the separatrix given by the barrier tops at $E=NJ$ the motion is oscillatory with time average $\langle \phi \rangle=0$ (plasma oscillations), but when the energy is above the separatrix the phase can continuously wind up in either the clockwise or anticlockwise directions. Because of the winding, the angular momentum also has a finite time-average implying that $\langle z \rangle \neq 0$ (macroscopic quantum self-trapping).

\subsection{Optomechanics}

The second system we consider  is the ``membrane-in-the-middle'' (MM) setup realized in
optomechanics experiments \cite{thompson08,jayich08}. It consists of an optical cavity divided in two by a partially transmissive and elastic membrane. The cavity is pumped by laser light through the end mirrors and the membrane is deformed by the radiation pressure upon it. The membrane can be pushed to the left or the right: if it is pushed to the right, say, it reduces the length of the right hand cavity and increases the length of the left hand cavity. This changes the resonance frequency for each cavity resulting in a change in the number of photons which in turn changes the radiation pressure (this feedback is the origin of the nonlinearity in this system). The total Hamiltonian is \cite{heinrich11}
\begin{equation}
\hat{H}_{\mathrm{MM}} = \hat{H}_{\mathrm{m}} +
\hat{H}_{\mathrm{l}} + \hat{H}_{\mathrm{int}} + \hat{H}_{\mathrm{p}}
\end{equation}
where 
\begin{eqnarray}
\hat{H}_{\mathrm{m}} &=& \frac{\hat{p}^2}{2m} + \frac{m
  \omega^2 \hat{x}^2}{2} \nonumber \\
\hat{H}_{\mathrm{l}} &=& g \left ( \hat{a}^\dagger_R \hat{a}_L +
  \hat{a}^\dagger_L \hat{a}_R \right )  \nonumber \\
\hat{H}_{\mathrm{int}} &=& \frac{2 \gamma}{\sqrt{V}} \hat{x} \hat{n} \nonumber \\
\hat{H}_{\mathrm{p}} &=& \eta_R \sqrt{V} \left ( \hat{a}^\dagger_R
  + \hat{a}_R \right ) + \eta_L \sqrt{V} \left ( \hat{a}^\dagger_L + \hat{a}_L \right ) \ ,
\end{eqnarray}  
are the Hamiltonians for the membrane (mechanical harmonic oscillator), light hopping between cavities by transmission through the membrane, radiation pressure,  and pump,
respectively.  Here, like in the previous example, the left- and
right-hand cavity modes are labeled by $L$ and $R$, respectively,
however, now these modes are occupied by photons instead of massive
particles. $V$ is the cavity mode volume and is related to the number
of photons in a cavity by $V = N/\rho$ where $\rho$ is the number
density of photons.  The parameters $\omega$ and $g$ are the natural oscillation frequencies of the membrane and light hopping, respectively, $\gamma$ gives the
interaction energy due to radiation pressure and $\eta_L$ and $\eta_R$
give the pumping strengths for the left and right cavities.  The relevant parameter in this system is $\Lambda_{\mathrm{MM}} =
(2g/m) ( 2 \gamma \eta/[\omega (g^2+ \kappa^2)] )^2$ where for $\Lambda_{\mathrm{MM}} > 1$ the ground state of the system
goes from being a centred membrane with an equal number of photons in
each cavity to a shifted membrane with a buildup of light in one
cavity over the other which is the result of breaking the
$\mathbb{Z}_2$ symmetry of the system.  

In experiments it is usually the case that the light field
evolves much faster than the membrane, i.e.\ $g \gg \omega$ \cite{bhattacharya08,larson11}, so
that the light `instantaneously' adjusts to the position of the membrane. The optical modes can then be adiabatically eliminated to give an effective
potential for the membrane alone.  To do this we assume the
light satisfies the stationary solutions of the equations of motion,
$\dot{\hat{a}}_R = \dot{\hat{a}}_L = 0$, giving
\begin{eqnarray}
\hat{a}^{\mathrm{s}}_R &=& - \frac{i \eta_R \kappa \sqrt{V}  + g \eta_L
  \sqrt{V} + \hat{x} \eta_R \gamma}{g^2+\kappa^2+\hat{x}^2 \gamma^2/V}
\nonumber \\
\hat{a}^{\mathrm{s}}_L &=& - \frac{i \eta_L \kappa \sqrt{V}  + g \eta_R
  \sqrt{V} - \hat{x} \eta_L \gamma}{g^2+\kappa^2+\hat{x}^2 \gamma^2/V}
\,  
\label{eq:ss}
\end{eqnarray}
where we have introduced a cavity decay rate $\kappa$.  We
obtain the effective potential by substituting Eqns.\ (\ref{eq:ss})
into 
\begin{equation}
\dot{\hat{p}} = - m \omega^2 \hat{x} - \frac{2 \gamma}{\sqrt{V}} \hat{n} = -
\frac{d V(\hat{x})}{d\hat{x}}
\end{equation}
which upon integration gives the effective Hamiltonian for the membrane \cite{mumford15}
\begin{equation}
\frac{\hat{H}^\prime_{\mathrm{MM}}}{V} = \frac{\hat{p}^2}{2 m} + \frac{m
  \omega^2 \hat{x}^2}{2} + \frac{4 g \eta^2}{g^2 + \kappa^2 +
  \hat{x}^2 \gamma^2}
  \label{eq:membraneHamiltonian}
\end{equation}
where the transformations $\hat{p} \rightarrow \hat{p} \sqrt{V}$ and
$\hat{x} \rightarrow \hat{x} \sqrt{V}$ were made, so in
$[\hat{x},\hat{p}] = i/V$ the limit $V
\rightarrow \infty$ is again the same as $\hbar \rightarrow 0$.  We
have also assumed the ground state is being pumped, which for $g >
0$ means $\eta_R = - \eta_L = \eta$ \cite{miladinovic11}. This Hamiltonian is of the desired form given by  Eq.\ (\ref{eq:ham}). Near the critical value of  $\Lambda_{\mathrm{MM}}$ it is sufficient to Taylor expand the effective potential up to quartic terms so that it can take on a double-well shape. The transition from a single- to a double-well describes the $\mathbb{Z}_2$ symmetry breaking transition where the membrane spontaneously displaces to the left or right. Furthermore, this is a dynamical phase transition as the cavity is pumped by laser light and hence is not in its ground state.

\subsection{Dicke model in the Holstein-Primakoff representation}

Lastly, we look at the Dicke model (DM) which describes a collection of
spin-1/2 particles coupled to a harmonic oscillator. In its original context this was used to model collective light emission (superradiance) by $N$ two-level atoms coupled to a single mode of
the electromagnetic field  \cite{dicke54}.  Unlike in the last example where we
eliminated the degrees of freedom of one part of the system, we keep both here. In a cold atom context the DM has been realized using a BEC inside an optical cavity \cite{baumann10,nagy10}, where the two `spin' states refer to two different translational modes of the atoms.
The DM Hamiltonian can be written
\begin{equation}
\hat{H}_{\mathrm{DM}} = \omega_0 \hat{S}_z + \omega \hat{b}^\dagger
\hat{b} + \frac{\chi}{\sqrt{2 S}} \left ( \hat{b}^\dagger + \hat{b}
\right ) \left ( \hat{S}_+ + \hat{S}_- \right )
\label{eq:DM}
\end{equation}
where the Schwinger representation has been used to describe the $N$
two-level systems, each with excitation frequency $\omega_0$,  as a large pseudospin of length $S = N/2$.  The electromagnetic field mode with frequency $\omega$ is acted on by the creation (annihilation) operator $\hat{b}^\dagger$
($\hat{b}$)  and the coupling with the spins is given
by $\chi$.  For $\Lambda_{\mathrm{DM}} = 2 \chi/\sqrt{\omega \omega_0}
> 1$ the ground state suffers a parity breaking ($\mathbb{Z}_2$) phase transition resulting in a spontaneous excitation of the harmonic oscillator, i.e.\ the coherent
emission of light by the atoms. The presence of external pumping of the cavity once again means that this is a dynamical rather than a ground state phase transition.  To describe the phase transition Emary and Brandes
\cite{emary03} used the Holstein-Primakoff representation  \cite{holstein49,ressayre75} of spin operators to write them in terms of ordinary annihilation and creation operators 
\begin{eqnarray}
\hat{S}_{+} = \hat{a}^\dagger \sqrt{2 S - \hat{a}^\dagger \hat{a}},
&& \hspace{5pt} \hat{S}_{-} = \sqrt{2 S - \hat{a}^\dagger \hat{a}} \,
\hat{a} \nonumber \\
\hat{S}_{z} = \hat{a}^\dagger \hat{a} - S &&
\end{eqnarray}
where $[\hat{a}, \hat{a}^\dagger] = 1$.  The Holstein-Primakoff representation is useful when the spin is only weakly excited above its ground state which is the extremal spin projection state $\vert  S, m=-S \rangle$, so that 
$\langle \hat{a}^\dagger \hat{a} \rangle /2 S \ll 1$, and the square
roots can be expanded in powers of $1/2S$.  By converting the annihilation and creation operators into position and momentum operators using the standard definitions
\begin{eqnarray}
\hat{b} \equiv \sqrt{\frac{\omega}{2}} \left ( \hat{x} +
  \frac{i}{\omega} \hat{p}_x \right ), && \hspace{10pt} \hat{b}^\dagger
\equiv \sqrt{\frac{\omega}{2}} \left ( \hat{x} - \frac{i}{\omega}
  \hat{p}_x \right ) \nonumber \\
\hat{a} \equiv \sqrt{\frac{\omega_0}{2}} \left ( \hat{y} +
  \frac{i}{\omega_0} \hat{p}_y \right ), && \hspace{10pt} \hat{a}^\dagger
\equiv \sqrt{\frac{\omega_0}{2}} \left ( \hat{y} - \frac{i}{\omega_0}
  \hat{p}_y \right ) \ , \nonumber \\
\end{eqnarray}
they were able to show that Eq.\ (\ref{eq:DM}) takes the form
\begin{eqnarray}
\hat{H}_{\mathrm{DM}} = && \frac{1}{2} \left ( \hat{p}_x^2 + \omega^2
  \hat{x}^2 + \hat{p}_y^2 + \omega_0^2 \hat{y}^2 \right ) \nonumber \\
&+& \chi \sqrt{\omega \omega_0} \hat{x} \biggl[ \left ( \hat{y} -
    \frac{i}{\omega_0} \hat{p}_y \right ) \sqrt{1 - \hat{\eta}}
  \nonumber \\ &+&
  \sqrt{1 - \hat{\eta}}  \left ( \hat{y} +
    \frac{i}{\omega_0} \hat{p}_y \right ) \biggr]
\label{eq:DM2}
\end{eqnarray}
where
\begin{equation}
\hat{\eta} = \left ( \omega_0^2 \hat{y}^2 + \hat{p}_y^2 - \omega_0
\right )/(4 S \omega_0) \, .
\end{equation}
Even though $\hat{H}_{\mathrm{DM}}$ has imaginary terms and a momentum
dependent potential, $V(\hat{x},\hat{y},\hat{p}_y)$, for $S \gg 1$ we
can approximate it by ignoring the commutation relation between the
operators in the square brackets.  With the transformations $\hat{x}
\rightarrow \hat{x} \sqrt{S}$, $\hat{p}_x \rightarrow
\hat{p}_x \sqrt{S}$, $\hat{y} \rightarrow \hat{y} \sqrt{S}$ and
$\hat{p}_y \rightarrow \hat{p}_y \sqrt{S}$, Eq.\ (\ref{eq:DM2}) becomes
\begin{eqnarray}
\frac{\hat{H}_{\mathrm{DM}}^\prime}{S} =  & \frac{1}{2} \left ( \hat{p}_x^2 + \omega^2
  \hat{x}^2 + \hat{p}_y^2 + \omega_0^2 \hat{y}^2 \right ) \nonumber \\
&+ 2 \chi \sqrt{\omega \omega_0} \hat{x} \hat{y} \sqrt{1 -
  \frac{\omega_0 \hat{y}^2}{4}} \, .
\end{eqnarray}
together with the now familiar commutation relations
$[\hat{x},\hat{p}_x] = \mathrm{i}/S$, $[\hat{y}, \hat{p}_y] = \mathrm{i}/S$
($[\hat{x},\hat{p}_y] = 0$ and $[\hat{y},\hat{p}_x ] =0$).  We can see that since we kept both parts of the system the Hamiltonian is two-dimensional and so generalizes the form given in Eq.\ (\ref{eq:ham}), but is nevertheless
of the form of kinetic plus potential terms and so our
proceeding analysis can be applied here as well.

In this section we have used various approximation methods to write the
Hamiltonians of some simple many-particle systems in the form of a single effective quantum
particle like in  Eq.\ (\ref{eq:ham}).  The results represent a semi-classical approach to
each system where we have assumed they are large enough to be approximated
by continuous spectra, but we do not take the thermodynamic limit, so
there are still canonical commutation relations to be obeyed.  It is
in this regime we will focus on investigating the quantum critical
nature of catastrophes. 

Once again, we emphasize that the catastrophes that exist in all three models described above properly live in Fock space. However, in the continuum approximation the fundamental discretization of Fock space vanishes, and hence the distinction between quantum catastrophes and the standard continuous wave catastrophes evaporates (for an analysis of a quantum catastrophe see \cite{mumford16}).  For simplicity, in the remainder of this paper we will keep coming back to the example
system of the two-site Bose-Hubbard model, although the basic results also apply to the optomechanical and Dicke models.

\begin{figure}[h]
		\begin{center}
			\includegraphics[width=1\columnwidth]{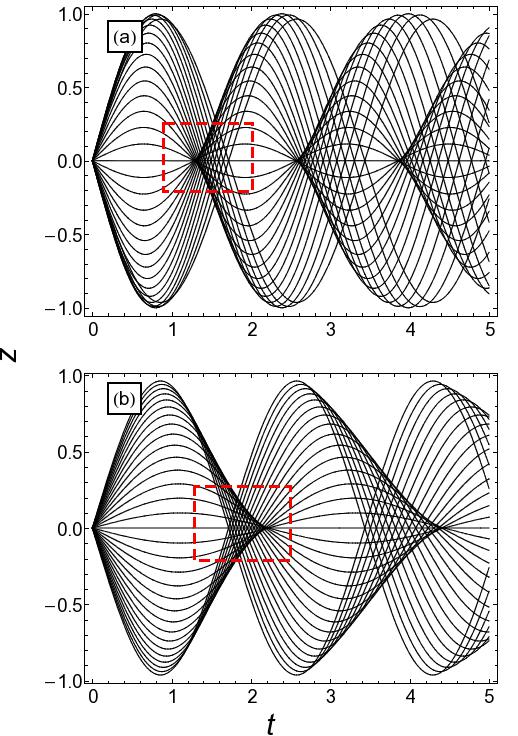}
		\end{center}
		\caption{Cusps in the classical dynamics of the two-mode Bose-Hubbard model. Each curve is a solution of the mean-field equations of motion (Hamilton's equations) and gives the number difference $z(t)$ between the left and right sites for $\Lambda_{\mathrm{BH}} = 0.5$.   The initial conditions are such that each trajectory starts with $z(0) = 0$ but has a different initial phase $\phi(0)$ sampled uniformly from $\phi(0) \in [0,2 \pi]$ in accordance with the truncated Wigner approximation. We have separated the trajectories into two groups: panel (a) shows those that oscillate around $\phi=0$ (plasma oscillations) and panel (b) shows those that oscillate around $\phi = \pi$ ($\pi$-oscillations). Both groups are excited under these conditions and we plot them separately for clarity. Near $z=\pm 1$ the cusps reach the maximum excitation possible in this system and hence curve off. This is a non-generic feature specific to the bounded Fock space of our system.   The red, dashed-boxed region indicates the approximate location of the generic or ``pure" cusp. Note that the quantum version of this figure is plotted in Figure \ref{fig:pigrid}(a). } 
		\label{fig:BJJ}
	\end{figure}

	\section{Catastrophes in Classical Dynamics}
\label{sec:cat}

In the truncated Wigner approximation (TWA) one attempts to mimic quantum dynamics by an ensemble of classical trajectories \cite{Sinatra2002,Polkovnikov2003}. This method has been implemented for the two-mode Bose-Hubbard model in \cite{Javanainen2013} where they also consider the effect of decoherence due to a continuous measurement of the number difference between the two sites, although we shall not include that additional feature here. The initial conditions for the classical trajectories are sampled from the initial quantum probability distribution, thus building in quantum fluctuations, but the subsequent time evolution of these trajectories is purely classical. 

For an initial state let us consider the physically realistic situation where two independent condensates with an equal number of atoms are suddenly placed in contact through a tunnelling barrier, i.e.\ a quench in the tunnelling rate from zero to a finite value specified by $\Lambda_{\mathrm{BH}}$. According to Heisenberg's uncertainty principle, if the number difference $z$ is exactly known then its conjugate variable $\phi$ is completely unspecified and hence the classical trajectories sampling the initial state all have $z(0)=0$ but differ in their initial value of the phase difference $\phi(0)$, being equally distributed over the range $(0,2 \pi]$. These trajectories are propagated in time by solving Hamilton's equations \cite{smerzi97}
\begin{eqnarray}
\dot{\phi} = \frac{\partial H_{\mathrm{BH}}}{\partial z} = \Lambda_{\mathrm{BH}} z +\frac{z}{\sqrt{1-z^2}} \cos \phi \\
 \dot{z} = - \frac{\partial H_{\mathrm{BH}}}{\partial \phi} = - \sqrt{1-z^2} \sin \phi \, 
\end{eqnarray}
obtained from the mean-field Hamiltonian given in Eq.\ (\ref{eq:BJJHam}).  The results are plotted in Figure \ref{fig:BJJ} for $\Lambda_{\mathrm{BH}} = 0.5$ where we see that a repeated series of cusp catastrophes are formed by the envelopes of the classical trajectories $z(t)$. To find the TWA (classical) prediction for the probability distribution in Fock space at time $t$ one should average over the trajectories, i.e.\  break the $z$ coordinate into little bins and count the number of trajectories that arrive in each bin. In this way one finds that the probability  diverges on the cusps as the number of trajectories becomes large (see, e.g., Figure 2 in \cite{odell12}).  It is worth pointing out that the cusps shown in Figure \ref{fig:BJJ} are not a special feature of the initial condition $z(0)=0$. Although this initial condition does give cusps which are symmetric about $z=0$, the structural stability of catastrophes ensures that they are robust to fluctuations in the initial conditions which can also be imbalanced (see Figure \ref{fig:posnegcusp} below).

The cusps arise from the focusing effect of the minima in the effective potential in the Hamiltonian. If the potential is replaced by its expansion up to second order around the origin, $V(\phi) \approx -1+[(1+\Lambda_{\mathrm{BH}})/2 \Lambda_{\mathrm{BH}}] \phi^2 $, the focusing becomes perfect due to the isochronous nature of harmonic potentials: each cusp is reduced to a single focal point. However, this is a non-generic situation because perfect focal points are unstable to perturbations such as the inclusion of the non-harmonic part of $V(\phi)$ which smears them out into cusps. The cusps are, by contrast, structurally stable. The cusps in Figure \ref{fig:BJJ} are also stable against changes to the initial conditions. These can be varied to include imbalanced wells, or take $z(0) \neq 0$. Under these changes the cusp is modified quantitatively but not qualitatively. It is also interesting to note that in the Bogoliubov theory for the weakly interacting Bose gas the equations of motion are linearized \cite{pitaevskii03}, meaning that $V(\phi)$ is replaced by its harmonic approximation, and hence the Bogoliubov theory is unsuitable for describing catastrophes in the two-mode problem.

\begin{figure}
\begin{center}
\includegraphics[width=1\columnwidth]{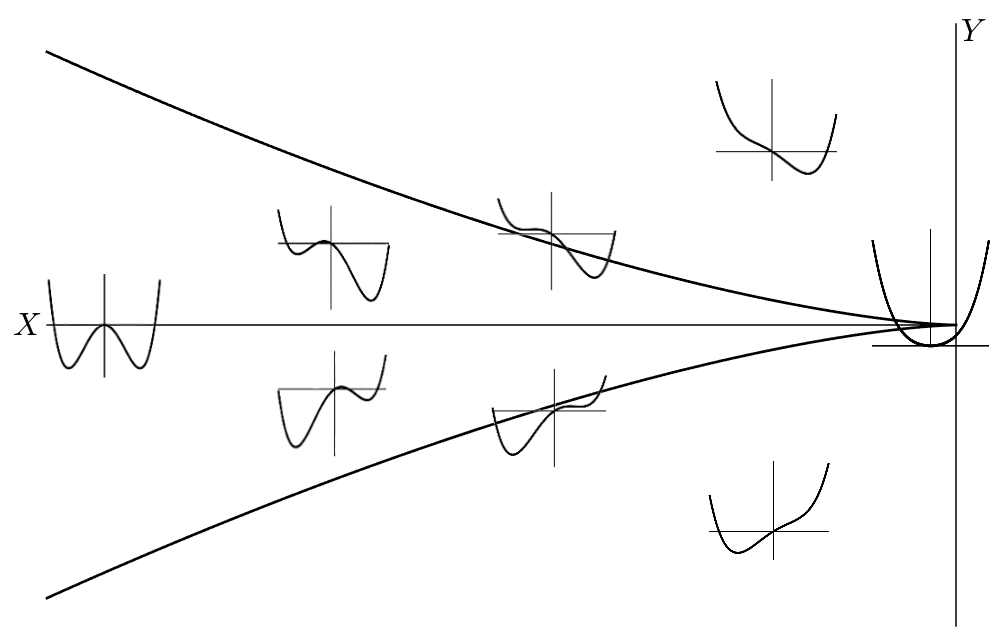}
\end{center}
	\caption{A plot of the canonical cusp as given by Eq.\ (\ref{eq:cusp}). It consists of two fold lines that meet at a cusp point. The insets at different points $(X,Y)$ show the cusp generating function $\Phi(s;X,Y) = s^4+Xs^2+ Ys$ plotted as a function of $s$. Each extremum of $\Phi$ corresponds to a classical trajectory; there are three at each point inside the cusp and one at each point outside. Note that Eq.\ (\ref{eq:cusp}) only has real solutions when $X$ is negative. By changing the signs of the terms in $\Phi(s;X,Y)$ the cusp can instead be made to live in the positive-$X$ half plane.} 
	\label{fig:actioncusp}
\end{figure}

To understand why we specifically see cusps in the two-dimensional ($z,t$) control space, consider the generating function/action $\Phi=s^4+X s^2+Y s$ for co-dimension 2 catastrophes in Table \ref{tab:catastrophes}. According to Hamilton's principle the classical trajectories are those for which the action is stationary with respect to variations in the state variables which characterize them. This gives  $\partial \Phi/\partial s =4s^3+2Xs+Y= 0$. On a catastrophe the action is stationary to higher order $\partial^2 \Phi/\partial s^2 = 12 s^2+2 X =0$;  physically this is the focusing condition. Eliminating $s$ from these two equations gives the equation for a cusp
\begin{equation}
Y = \pm \sqrt{\frac{8}{27}} (-X)^{3/2} 
\label{eq:cusp}
\end{equation}
and is plotted in Figure \ref{fig:actioncusp}. The insets at different points $(X,Y)$ depict the action $\Phi(s;X,Y)$ as a function of $s$. Being a quartic function, $\Phi$ has at most three stationary points; each stationary point corresponds to a classical trajectory. We see that there are three classical trajectories reaching each point inside the cusp and just one reaching each point outside. As we cross one of the edges of the cusp (known as fold lines) two of the solutions coalesce and annihilate leading to a singularity. However, the most singular part is the point of the cusp where all three solutions coalesce at once. In a specific system the canonical coordinates $\{ X,Y \}$ will not generally correspond to the actual physical coordinates, but transformations can (in principle) be found that relate the two. We will see an example of this in Section \ref{sec:quantum}.

\begin{figure*}[!ht]
\begin{center}
\includegraphics[width=2.0\columnwidth]{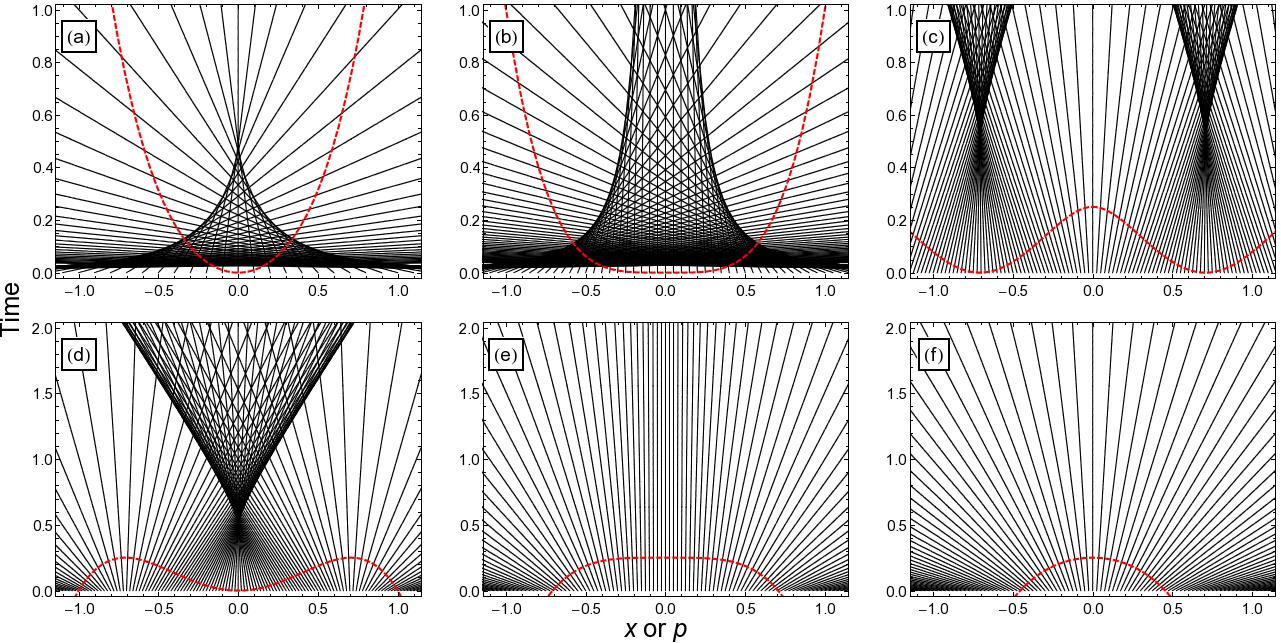}
\end{center}
	\caption{Classical dynamics for kicked Hamiltonians with a quartic potential whose shape is indicated by the red dashed curve in each panel. The top row has $a_{4}>0$ and the bottom row has $a_{4}<0$.  The reduced parameter $\lambda$ decreases from left to right so that $a_2 = 2$, $a_2  = 0$ and $a_2 = -2$ in the left hand, central and right hand columns, respectively.  Each black solid line is a classical trajectory with a different initial $x \in [-3,3]$. If the kinetic term is kicked then the dynamics take place in the $(p,t)$-plane, whereas if the potential term is kicked they take place in the $(x,t)$-plane.  The images are similar to what one would find in the geometric theory of light where incoming parallel rays (not shown) reflect from a mirror with the same local curvature as the potential.}
\label{fig:cuspkickgrid}
\end{figure*}

Structural stability implies that we need not be concerned with the exact shape of the potential but rather with its general features such as the number of stationary points. Accordingly, in the rest of this paper we will confine our attention to a general quartic potential 
\begin{equation}
V(x) = a_0 + a_2 x^2 + a_3 x^3 + a_4 x^4 \ .
\label{eq:approxpot}
\end{equation}
In general the coefficients $a_2$ and $a_4$ depend on the parameters
of the system.  If we assume there is one such parameter $\Lambda$
(like the ones identified in each example in Section \ref{sec:approx})
which drives the system through a second order phase transition
then we can take inspiration from the Landau theory of continuous phase transitions and approximate the coefficients near the critical point at $\Lambda_c$ as $a_2(\Lambda)
\approx \lambda/2$ and $a_4(\Lambda) \approx \pm 1$,
where $\lambda = (\Lambda - \Lambda_c)/\Lambda_c$ is the reduced
driving parameter.  We have set $a_0 = 0$ without loss of generality
because this just results in an overall shift of the energy.   On the one hand, when $a_4 > 0$ (with $a_3 = 0$) we have either a single- or double-well potential depending upon whether $\lambda$ is positive or negative. On the other hand, when $a_4 < 0$ (with $a_3 = 0$) for $\lambda >0$ there is a local minimum at $x = 0$ sandwiched between two global maxima
at $x_{\pm} = \pm \sqrt{\lambda}/2$, and for $\lambda <0$
there is a global maximum at $x = 0$.  This latter situation describes, for example,  $\pi$-oscillations providing the quartic potential is understood as a Taylor series expansion about the point $\phi=\pi$. At the critical point $\lambda_c = 0$, dynamics near this region become unstable resulting in
exponential divergence away from it.  This is important for
the fate of $\pi$-oscillation cusps because when the phase transition occurs the potential around
$x = 0$ no longer focuses trajectories but instead becomes an unstable stationary point that defocuses and destroys the cusps.

\section{$\delta$-kicked Hamiltonians}
\label{sec:kickedHamiltonians}

A further simplification we shall make at this point is to consider $\delta$-kicked Hamiltonians. $\delta$-kicks play an important role in molecular physics where trains of short laser pulses are used to align molecules \cite{leibscher03,dion02,pruna02} and in experiments involving a small number of pulses molecules have been shown to exhibit ``classical alignment echoes'' where the initial alignment is revived after initially collapsing \cite{karras15}. We note that in the kicked rotor problem it is known that a cusp can form in the angular position distribution \cite{leibscher04} and also in the angular momentum distribution \cite{floss15}. 
In cold atom experiments one can exert real-time control over both the trapping potential and the interactions between the atoms which allows for a broad range of options for kicking the system into a non-equilibrium state. For example, the $\delta$-kicked rotor can be realized in a cold atomic gas by flashing on and off an optical lattice \cite{moore95}, and in the case of a three-frequency periodic $\delta$-kick the system displays a form of Anderson localisation in time \cite{casati89} at a critical kicking strength (equivalent to disorder strength). The Green's function for the probability distribution in this case happens to be an Airy function which gives it a scaling invariance characteristic of a second-order phase transition \cite{lemarie10}.  The critical behaviour of the $\delta$-kicked Lipkin-Meshkov-Glick model has been investigated in reference \cite{bastidas14}.

We shall consider the simplest case of a single $\delta$-kick to one of the terms in the Hamiltonian while the rest is held constant. This type of time evolution facilitates analytical results and allows a very clean realization of the canonical wave catastrophes.  In fact, one can kick either of the terms in the Hamiltonian (\ref{eq:ham}) as what really counts is that we have two non-commuting pieces at some instant, one of which is also non-linear. Thus, we consider two cases 
\begin{eqnarray}
\hat{H}_1 &=&  \delta(t) \frac{\hat{p}^2}{2} + V(\hat{x}) \quad \mbox{\textit{Case 1}} \\
\hat{H}_2 &=& \frac{\hat{p}^2}{2} + \delta (t)  V(\hat{x}) \quad \mbox{\textit{Case 2}} \, ,
\end{eqnarray} 
where for now we have set $N$ to unity.  After the kick the system evolves due to only one term which makes an analytical description easier, especially in the classical case where Hamilton's equations $\dot{x} = \partial H / \partial p$, and $ \dot{p} = - \partial H / \partial x$
can be solved trivially.  For $\hat{H}_1$ one finds 
\begin{equation}
x(t) = x(0) = p_0, \hspace{15pt} p(t) = t \, F(x(0))+ p_0,
\end{equation}
 and for $\hat{H}_2$ 
\begin{equation}
x(t) = t \, p(0) + x_0 , \hspace{15pt} p(t) = p(0) = F(x(0)).
\end{equation}
  In both expressions $F(x) = -\partial V/\partial x$ is the force.  We therefore see that the classical trajectories are straight lines in either the $(p,t)$- or $(x,t)$-plane with slopes determined by the initial force or momentum.

\begin{figure}[!h]
\begin{center}
\includegraphics[width=0.75\columnwidth]{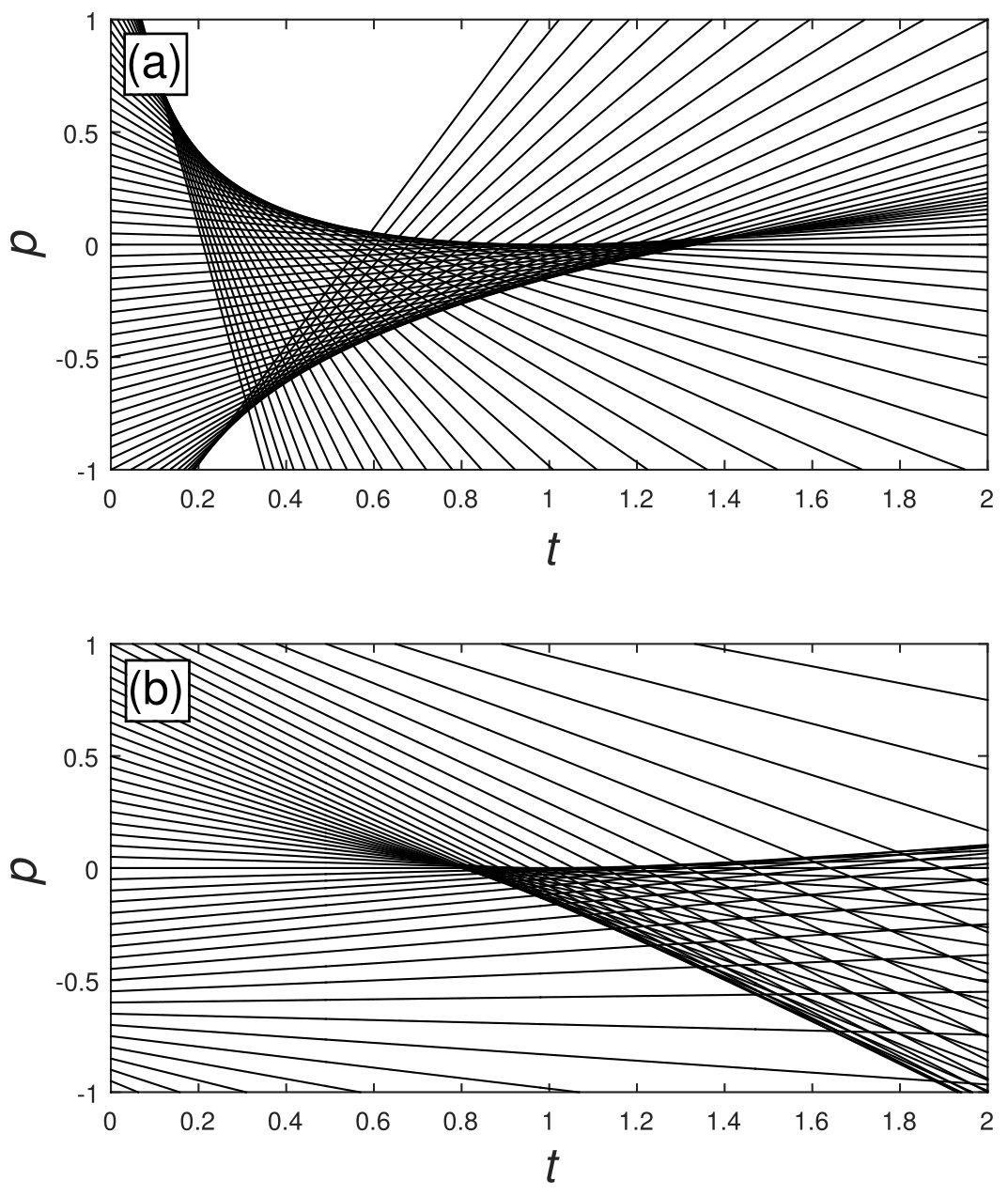}
\end{center}
	\caption{Trajectories for the case of a kicked kinetic term ($H_{1}$) with different
          initial $p \in [-2,2]$ for different values of $a_4$: (a)
          $a_4 = 1$ and (b) $a_4 = -1$.  For both images $a_2 = a_3 = 1$.} 
	\label{fig:posnegcusp}
\end{figure}

The classical trajectories following a kick for various incarnations of the quartic potential are plotted in Figure \ref{fig:cuspkickgrid}. In the top row $a_4>0$, and as the potential turns from a single to a double well the dynamics evolve from featuring a single cusp to two cusps. Note that the new cusps open in the opposite direction to the original one.  At the transition point at $\lambda = 0$ the cusp point is pushed off to $t= \infty$, a feature that may be viewed as an example of critical slowing down of the dynamics. Similarly, the two new cusps start at $t=\infty$ at the transition point and are brought down to finite times past the transition. In the bottom row $a_4<0$, and there is a single cusp generated by the central minimum of the potential when  $\lambda > 0$, which becomes a maximum for $\lambda< 0$ leading to  a divergence of the trajectories. The difference between positive and negative $a_4$ is also shown in Figure \ref{fig:posnegcusp}, as well as including the effect of an asymmetric potential by having $a_3 \neq 0$.  We see that the images still retain their qualitative cusp form, but are now skewed by the asymmetry.

\section{Catastrophes in Quantum Dynamics}
\label{sec:quantum}

\subsection{Mapping to the Pearcey function}

In the quantum description of the kicked system the evolution operator can be written as the product of two terms; one describing the kick at $t=0$ and the other describing the subsequent evolution \cite{haake09}. As for the classical problem, we will consider two cases; {\it Case 1:} Hamiltonians with a kicked kinetic term ($H_{1}$), and {\it Case 2:}  Hamiltonians with a kicked potential term ($H_{2}$). The evolution operators in these two cases are $\mathcal{\hat{U}}_1 = \rme^{-\rmi
  V(\hat{x}) t} \rme^{-\rmi\hat{p}^2 /2}$ and $\mathcal{\hat{U}}_2 =
\rme^{-\rmi \hat{p}^2t/2} \rme^{-\rmi V(\hat{x}) }$, respectively.  The stability of
the cusp to perturbations allows us to choose a wide range of initial
states, however, with simplicity in mind we choose the 
ground state of the non-kicked term in the $\lambda > 0$ (symmetric) phase,
so for case one $\vert \psi_0 \rangle_1 = \vert x = 0 \rangle =
\int_{-\infty}^{\infty} \vert p_0 \rangle \rmd p_0$ and for case two
$\vert \psi_0 \rangle_2 = \vert p = 0 \rangle  =
\int_{-\infty}^{\infty} \vert x_0 \rangle \rmd x_0$.  Applying the
evolution operators to these initial states gives the
amplitude of being at any point in $x$ or $p$ at time $t$  
\begin{eqnarray}
\psi_1(p,t) &=& \int^\infty_{-\infty} \rmd p_0 \langle p \vert
\mathcal{\hat{U}}_1 \vert p_0 \rangle \nonumber \\
&=& \frac{\rme^{- \rmi \pi/4}}{\sqrt{2
      \pi}} \int^\infty_{-\infty} \rmd x_0 \, \rme^{\rmi \left [ \frac{x_0^2}{2} - px_0 -
      V(x_0)t  \right ]} \label{eq:wf1} \\
\psi_2(x,t) &=& \int^\infty_{-\infty} \rmd x_0 \langle x \vert
\mathcal{\hat{U}}_2 \vert x_0 \rangle \nonumber \\
&=& \frac{\rme^{-\rmi \pi/4}}{\sqrt{2
    \pi t}} \int^\infty_{-\infty} \rmd x_0 \, \rme^{\rmi \left [ \frac{(x-x_0)^2}{2t}
  - V(x_0) \right ]} \, . \label{eq:wf2}
\end{eqnarray}
To make the connection to CT we substitute the quartic potential defined in Eq.\ (\ref{eq:approxpot}) into Eqns.\ (\ref{eq:wf1}) and (\ref{eq:wf2}) giving
\begin{eqnarray}
\psi_1(p,t) &=& \frac{\rme^{\rmi \theta_1 (p,t;\mathbf{a})}}{\sqrt{2
    \pi \sqrt{a_4 t}}}
\mathrm{Pe}[X_1(t;\mathbf{a}),Y_1(p,t;\mathbf{a})] \label{eq:Pe1} \\
\psi_2(x,t) &=& \frac{\rme^{\rmi \theta_2 (x,t;\mathbf{a})}}{\sqrt{ 2
    \pi t \sqrt{a _4}}}
\mathrm{Pe}[X_2(t;\mathbf{a}),Y_2(x,t;\mathbf{a})] \,  \label{eq:Pe2}
\end{eqnarray}	
where $\mathrm{Pe}[X,Y]$ is the Pearcey function \cite{pearcey46}
\begin{equation}
\mathrm{Pe}[X,Y] = \int_{-\infty}^{\infty} \rmd s \, \rme^{-\rmi (Y
  s+Xs^2+s^4)} \ ,
\label{eq:pearcey}
\end{equation}
which is the wave catastrophe corresponding to the cusp \cite{berry80,pearcey46,NISThandbook,connor73,berry79,stamnes83,kaminski99} and is plotted in Figure \ref{fig:quantcusps}.  The phase factors multiplying the wave functions are 
\begin{align}
	\theta_1(p,t;\mathbf{a}) =&\;
        a_0  t- \frac{a_3^2}{32 a_4^2} - \frac{3a_3^4 t}{256 a_4^3} +
        \frac{a_2 a_3^2 t}{16 a_4^2} + p x_m+\frac{x_m^2}{2} \\
	\theta_2(x,t;\mathbf{a})
        =&\;a_0 - \frac{a_3^2}{32 a_4^2 t} - \frac{3 a_3^4}{256 a_4^3}
        + \frac{a_2 a_3^2}{16 a_4^2} - \frac{a_3 x}{4 a_4 t} - \frac{x^2}{2t}
	\label{eq:C}
	\end{align}
where $\mathbf{a}=\{a_{0},a_{1},a_{2},a_{3}\}$ are the four parameters specifying the quartic potential. The quantity $x_{m}$ is required if the quartic potential $V(x)$ is a Taylor series expansion about the point $x_{m} \neq 0$ in which case all values of $x$ are measured from $x_{m}$; otherwise $x_{m}=0$. The transformation between the physical coordinates and parameters and the canonical state variables and control parameters is given, for {\it Case 2}, by	
\begin{eqnarray}
s_{2} = a_4^{1/4} \left (x_0 +\frac{a_3}{4 a_4} \right )  \nonumber \\
X_2(t;\mathbf{a}) = -\frac{\left(3 a_3^2 t+a_4 (4-8 a_2 t)\right)}{8 a_4^{3/2} t} \nonumber \\
Y_2(x,t;\mathbf{a}) = \frac{ \left(8 a_4^2 x+a_3^3 t+a_3 a_4 (2-4 a_2 t)\right)}{8 a_4^{9/4} t}.
\label{eq:parameters}
\end{eqnarray}
We see that classical paths, as characterized by $s_{2}$, are specified by their initial $x$ coordinate $x_{0}$. Also, the canonical control parameter $Y$ mixes the physical coordinates $(x,t)$ whereas $X$ is a function purely of $t$. For {\it Case 1} the transformations are closely related to those of {\it Case 2}
\begin{eqnarray}
s_1 = t^{1/4} s_2 \nonumber \\
X_1(t;\mathbf{a}) = \sqrt{t} X_2(t;\mathbf{a}) \nonumber \\
Y_1(p,t;\mathbf{a}) = t^{3/4} Y_2(p,t;\mathbf{a}) \ .
\label{eq:parameters1}
\end{eqnarray}

It is easier to see the fine details
within a cusp opening in the positive $t$ direction than the negative 
$t$ direction, so we will assume $a_4 <0$.  With our definitions of
$a_2$, $a_3$ and $a_4$ in the previous section, Eq.\ (\ref{eq:parameters}) becomes
\begin{eqnarray}
s_2 = x_0   \nonumber \\
X_2(t;a) = \frac{(1- \lambda t)}{2t} \nonumber \\
Y_2(x,t) = \frac{x}{t} \, ,
\label{eq:parameters2}
\end{eqnarray}
where the relation of the variables between the two cases is the same
as those given in Eq.\ (\ref{eq:parameters1}).

\subsection{Scaling exponents}

\begin{table}
\caption{\label{tab:CE} Critical scaling exponents of $\lambda$ for the cusp
  catastrophe, where the critical point is at $\lambda=0$. The first two columns refer to classical properties of the cusp: the exponent for the cusp size refers to the scaling in the transverse direction ($p$ or $x$), and the exponent for the position refers to the location of the cusp point $t_{\mathrm{cusp}}$ in the time direction. The remaining three columns refer to quantum properties and assume that the classical properties of the cusp are held fixed by working in the ($\zeta,\tau$) coordinates. The exponent for the probability density at the cusp point $\vert \psi_{\alpha} (0,1,\lambda) \vert^2$ is $2 \alpha \beta$, where $\beta$ is the Arnold index and $\alpha=1,2$ for {\it Case 1,2}. The last two columns give the scaling of the interference fringes, where $\sigma_{X}$ and $\sigma_{Y}$ are the Berry indices. Thus, as $\lambda \rightarrow 0$ the fringe spacing diverges as $\lambda^{-\alpha \sigma_{X}}$ and $\lambda^{-\alpha \sigma_{Y}}$ in the $X$ and $Y$ directions, respectively.}
  \begin{indented}
\item[] \begin{tabular}{@{} llllll} 
  \br   
  & \multicolumn{2}{l}{CLASSICAL} & \multicolumn{3}{l}{QUANTUM} \\   
  &  \crule{5}  \\           
  Kicked Term & Size & Position & $\vert \psi \vert^2$ & 
  $\alpha \, \sigma_X$ & $\alpha \, \sigma_Y$  \\
  \mr
  Kinetic (\textit{Case 1}) & 1/2 & -1 & 1/2  & 1/2 & 3/4  \\ 
  Potential (\textit{Case 2}) & 1/2 & -1 & 1  & 1 & 3/2  \\
  \br
\end{tabular}
\end{indented}
\end{table}

The critical behaviour of the ground states of the models studied in this paper have been investigated by a number of authors. For example, the critical exponents for the two-mode Bose-Hubbard model have been calculated in reference \cite{buonsante12}, and for the closely related Lipkin-Meshkov-Glick model in references \cite{dusuel04,dusuel05}. Similarly, the critical exponents of the Dicke model have been investigated in references \cite{emary03,vidal06}. Part of the power of the methods developed in this paper is that they give us access to the scaling properties of non-equilibrium states. In particular, in the vicinity of a catastrophe the quantum wave function obeys a remarkable self-similarity relation given in Eq.\ (\ref{eq:wfscaled}) below, with respect to the scale factor $\lambda$ and this allows us to quantify the non-equlibrium critical behaviour in terms of critical exponents.   Consider first the classical scaling which governs both the position and size of the cusp.  From Fig.\ \ref{fig:actioncusp} we see that the cusp point is located at $(X_{\mathrm{cusp}} =0, Y_{\mathrm{cusp}} = 0 )$ in the canonical coordinates.  Using Eq.\ (\ref{eq:parameters2}) to convert to physical coordinates we find that the cusp point is shifted to finite times $( t_{\mathrm{cusp}} = \lambda^{-1},x_{\mathrm{cusp}}=0 )$.
One can think of $t_{\mathrm{cusp}}$ as the time it takes the system
to respond to the initial kick: the fact that $t_{\mathrm{cusp}} \rightarrow \infty$ as $\lambda \rightarrow 0$ can be viewed as critical slowing down. Using $t_{\mathrm{cusp}}$ as the natural time scale allows us to define a time coordinate $\tau= t/t_{\mathrm{cusp}} = \lambda t$ which is invariant with $\lambda$. The analogous coordinate for the transverse direction is obtained by substituting $X$ and $Y$ in the canonical cusp formula given in Eq.\ (\ref{eq:cusp}) by the relevant quantities according to the above transformations and then replacing the time coordinate by the scaled time $\tau=\lambda t$. One finds  $p \propto \lambda^{1/2}$ and  $x \propto \lambda^{1/2}$ for {\it Case 1} and {\it Case 2}, respectively.  Thus, as the critical point is approached the cusp not only starts at later times but also shrinks in its transverse extent. An invariant coordinate for the transverse direction can therefore be defined as  
$\zeta =  x / \lambda^{1/2} = p/  \lambda^{1/2}$.

The quantum case is richer than the classical one due to the interference pattern decorating the cusp, as shown in Figure \ref{fig:quantcusps}. To get at the purely quantum features we work in the $(\zeta,\tau)$ coordinate system because these make Hamilton's equations invariant with $\lambda$ so that the cusp remains fixed in the $(\zeta,\tau)$-plane even as $\lambda$ is varied. Crucially,  the action is not scale
invariant and this is the source of the extra scaling properties of the quantum problem. Substituting in the new variables gives $\Phi_\alpha
\to \lambda^{\alpha} \Phi_\alpha$, where $\alpha = 1,2$ for {\it Case 1}
and {\it Case 2}, respectively.  The index $\alpha$ has no physical
significance and is only used for convenience in distinguishing the
different scalings between the two cases.  The factor of $\lambda^{\alpha}$ does not appear in the generating function for the canonical Pearcey function, but it can be absorbed into the control parameters and state variable if they are rescaled in a particular way that depends on three indices: $\beta$, $\sigma_{X}$ and $\sigma_{Y}$. The first index is known as the Arnold index, and the other two as  Berry indices. The rescaling thus returns us to the Pearcey function but with new control parameters scaled by $\lambda^{\alpha}$ and forms the basis for identifying the scaling properties of the catastrophe as $\lambda$ is varied.   Following this procedure through, we find we can write the wave functions in Eqns.\ (\ref{eq:Pe1}) and (\ref{eq:Pe2}) in the manifestly 
self-similar form
\begin{equation}
\psi_\alpha (\zeta, \tau; \lambda) \propto \left ( \frac{\lambda}{\tau}
\right )^{\alpha \beta} \mathrm{Pe} \left [ \lambda^{\alpha \, \sigma_X}
  X_\alpha, \lambda^{\alpha \, \sigma_Y} Y_\alpha  \right ]
\label{eq:wfscaled}
\end{equation}
where the proportionality sign indicates that we have neglected
overall phase and constant factors as they play no role in the
following analysis.  A derivation of Eq.\ (\ref{eq:wfscaled}) for {\it
  Case 2} is given in  \ref{app:derivation}.  The Arnold index governs how the amplitude of the wave function depends on the scale factor $\lambda$. In the case of the cusp catastrophe it takes the value $\beta = 1/4$ \cite{berry80}. The Berry indices dictate how rapidly the interference pattern varies in control space: in general the scaling in each direction is different and for the cusp they are $\sigma_X =1/2$ and  $\sigma_Y = 3/4$ \cite{berry80}.

With the wave function in the form of Eq.\ (\ref{eq:wfscaled}) it is easy to see that the probability density in Fock space at the cusp point scales as $\vert \psi_\alpha \vert^2 \propto \lambda^{2 \alpha \beta}$, and so for the two cases we have $\vert \psi_1 \vert^{2} \propto
\lambda^{1/2}$ and $\vert \psi_2 \vert^2 \propto \lambda$.  Thus, as $\lambda \rightarrow 0$ the cusp melts away, which is expected since the focusing region of the effective potential
shrinks (when $a_{4}<0$) causing fewer Fock states to contribute to the cusp. The interference pattern, meanwhile, varies more slowly as $\lambda \rightarrow 0$ with the fringe spacing tending to infinity in this limit. The scaling properties of the cusp wave function are summarized in Table \ref{tab:CE}.

So far we have set $N$ to unity, but now we will take a look at the
effects of its inclusion.  In each example we gave in Sec.\ \ref{sec:approx} 
we saw that the transformations made to the original many-particle Hamiltonian converted it to an effective single particle Hamiltonian $\hat{H} \to N \hat{H}^\prime$. The action 
undergoes the same transformation and this implies that the Pearcey
function changes to $\int_{- \infty}^{\infty} \rmd s \, \rme^{-\rmi
  N \left( Ys + Xs^2 + s^4 \right )}$, which means that $1/N$ plays the same role as $\hbar$ does in single particle path integrals. In particular, the thermodynamic limit, $N
\to \infty$ is the same as the classical limit, $\hbar \to 0$.  Furthermore, we see that $N$ multiplies the action in the same way as $\lambda^{\alpha}$ did above, and thus $\lambda^{\alpha}$ is replaced by $\lambda^{\alpha} N$ in the full theory. This implies that there is a clash of limits between the thermodynamic limit $N \rightarrow \infty$ and the `critical' point   $\lambda \rightarrow 0$.

\subsection{Vortices in Fock space + time}

\begin{figure}
\begin{center}
\includegraphics[width=1\columnwidth]{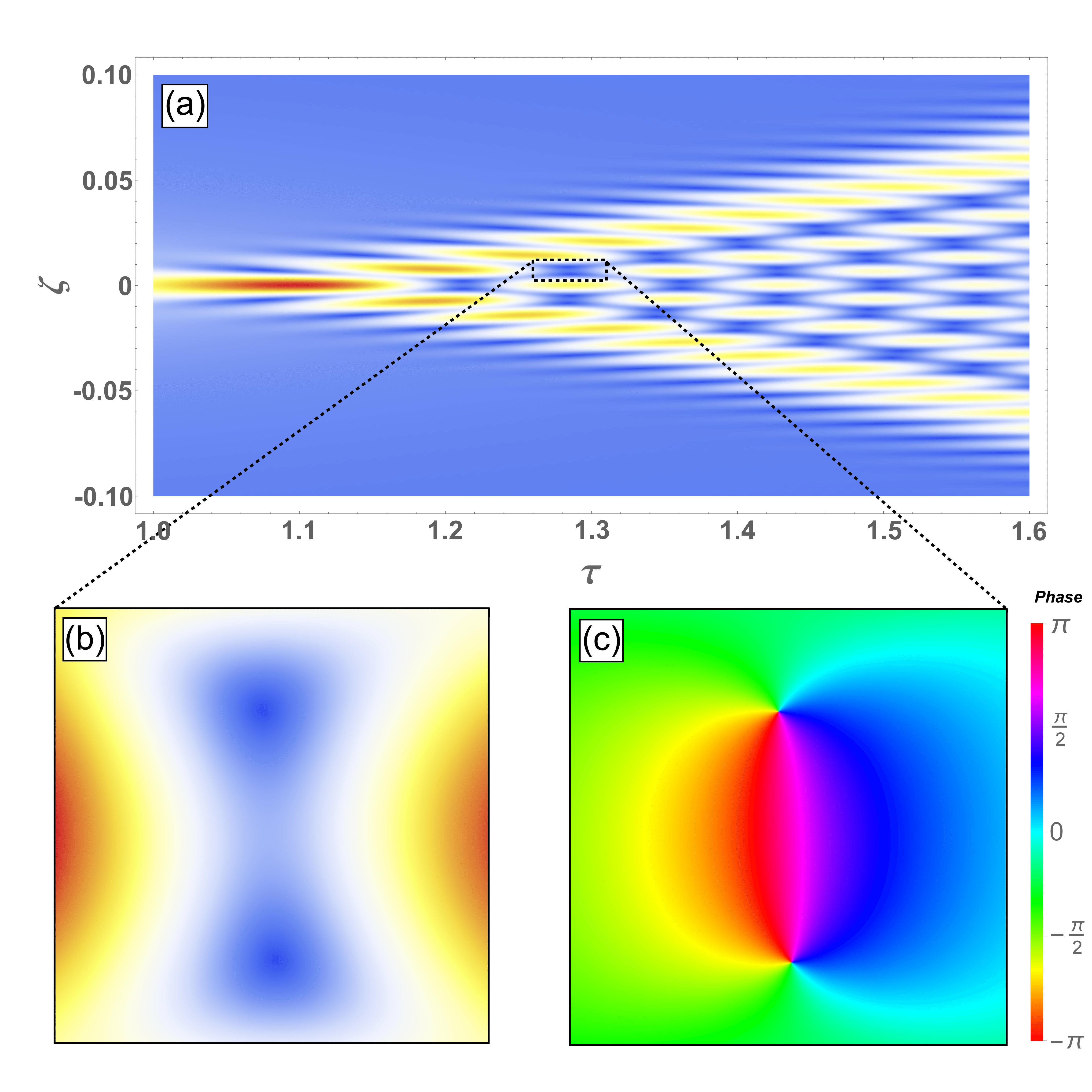}
\end{center}
	\caption{The wave function of a cusp catastrophe (the Pearcey function) for $N=1$ plotted as a function of the scaled control parameters $\zeta$ and $\tau$, i.e.\ Fock space and time. One sees that the underlying classical skeleton provided by the cusp is decorated by an intricate interference pattern with several levels of structure, and a slice at a fixed time through a fold line gives an Airy function provided one is not too close to the cusp point. In the ($\zeta,\tau$) coordinate system the classical cusp structure is held fixed, but the interference pattern evolves with $\lambda^{\alpha} N$. In the deep quantum regime where $\lambda^{\alpha} N \le 1$ the `fringes' are large. By contrast, in the opposite semiclassical regime $\lambda^{\alpha} N \gg 1$ the oscillations are very rapid and the fringe spacing is small.  Inside the cusp there is a network of vortex-antivortex pairs. Panels: (a) $\vert \psi_2\vert^2$,  (b) a closeup of a 
        vortex-antivortex pair which together form a dipole, and (c) the phase of the same pair as (b). In (a) and (b) blue indicates a small amplitude and red a large amplitude.} 
	\label{fig:quantcusps}
\end{figure}

Another remarkable feature of the interference pattern described by the Pearcey function is that it contains an intricate network of nodes \cite{pearcey46,berry79,kaminski99}. This `fine structure' can be seen by zooming in on the wave function as shown in panels (b) and (c) of Figure \ref{fig:quantcusps}. Examining the phase $\chi$ reveals that the nodes coincide with phase singularities where $\chi$ takes all possible values. Furthermore, $\chi$ circulates around the nodes in either a clockwise or anticlockwise sense  such that in going around once it changes by $\pm 2 \pi$,
\begin{equation}
\oint \rmd \chi = \pm 2 \pi \, .
\label{eq:phasechange}
\end{equation}
This is a topological feature that doesn't depend on the path of
integration providing it only encircles one node. All these properties
are familiar from quantized vortices that occur in coordinate space in
superfluids, type II superconductors and also optical fields (where
they are referred to as dislocations \cite{berry80,nye99}). The
difference is that here they occur in Fock space plus time. Note that the phase of
the Fock space amplitudes should not be confused with, e.g.\ the
relative phase in the two-mode Bose-Hubbard model, which is a different object.

\begin{figure}
	\begin{center}
		\includegraphics[width=1\columnwidth]{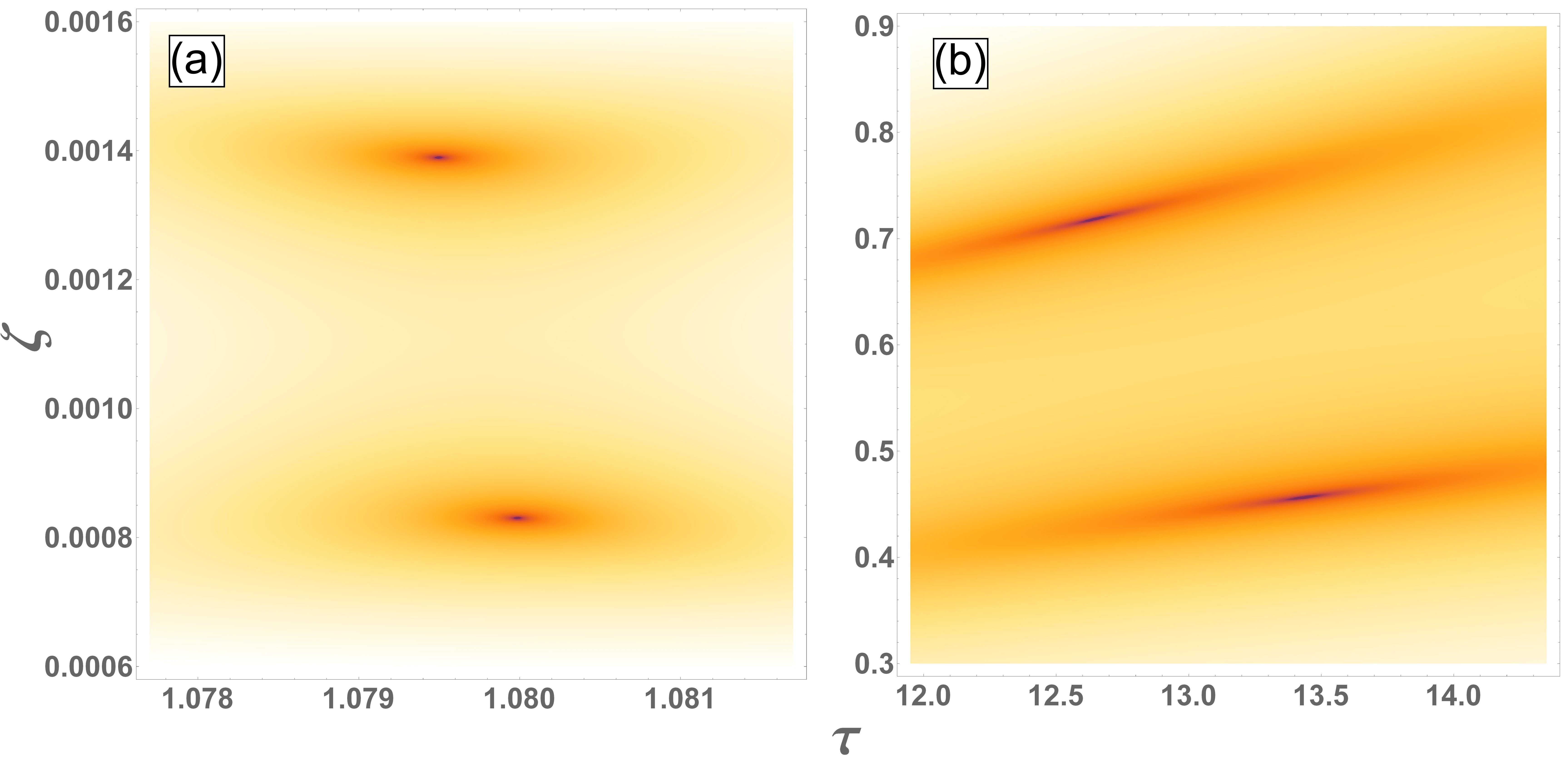}
	\end{center}
	\caption{Evolution of a vortex-antivortex pair as $\lambda$ is varied. Each panel shows $|\psi_2|$ plotted in Fock space plus time in the immediate vicinity of the same pair of vortices for: (a) $\lambda=150$ (far from the phase transition), and (b) $\lambda=12$ (approaching the phase transition). In order to demonstrate the apparent stretching of the vortices as they move apart, the aspect ratio $\Delta\tau/\Delta\zeta$ for the window remains unchanged.} 
	\label{fig:StretchedVorts}
\end{figure}

Inside the cusp the vortices are arranged in vortex-antivortex pairs, whereas outside the cusp there is a line of single vortices along each fold line. The Berry indices govern the scaling of distances in the control plane and so can tell us how the separation between a vortex and its antivortex changes with $\lambda$.  For a vortex-antivortex pair at positions
$(X_\alpha^{\mathrm{v}},Y_\alpha^{\mathrm{v}})$ and
$(X_\alpha^{\mathrm{av}},Y_\alpha^{\mathrm{av}})$, respectively, the
physical distance between them, $d_\alpha$, scales as 
\begin{equation}
d_\alpha = \sqrt{\frac{(X_\alpha^\mathrm{v} -
    X_\alpha^\mathrm{av})^2}{\lambda^{2 \alpha \sigma_X}}+
  \frac{(Y_\alpha^\mathrm{v} - Y_\alpha^\mathrm{av})^2}{\lambda^{2
      \alpha \sigma_Y}}} \, ,
\end{equation}
and so increases as $\lambda_c$ is
approached.  However, since $\sigma_X \neq \sigma_Y$ the two
directions do not scale in the same way and the vortices become
stretched out anisotropically. This effect persists in the $(\zeta,\tau)$ coordinates as shown in Figure \ref{fig:StretchedVorts}.

The scaling of distances in the classically invariant $(\zeta, \tau)$-plane is less obvious
because $\zeta$ and $\tau$ are functions of $X$ and $Y$. However, we can get the leading order behavior as $\lambda
\rightarrow 0$.  First, we note a given vortex moves around within the
cusp as $\lambda$ is varied such that $\lambda^{\alpha \sigma_X} X_\alpha$ and
$\lambda^{\alpha \sigma_Y} Y_\alpha$ remain constant.  If we find a
particular vortex for a given $\lambda$ such that $\lambda^{\alpha
  \sigma_X} X_\alpha = A_\alpha$ and $\lambda^{\alpha \sigma_Y}
Y_\alpha = B_\alpha$ where $A_\alpha <0$ and $B_\alpha$ are constants,
then we can find out how the vortices scale in $\zeta$ and $\tau$.
Using Eqns.\ (\ref{eq:parameters1}) together with Eqns.\
(\ref{eq:parameters2}) we find for {\it Case 1}
\begin{equation}
\tau - 1 = \frac{2 A_1^2}{\lambda^{2 \sigma_X}} \left ( 1 +
  \sqrt{1+\frac{\lambda^{2 \sigma_X}}{A_1^2} }\right ) \, ,
\end{equation}
so for $\lambda^{\sigma_X} \ll A_1$ we have $\tau \propto
\lambda^{-2 \sigma_X} = \lambda^{-1}$ and therefore $\zeta \propto
\lambda^{-\sigma_Y - 1/4} = \lambda^{-1}$.  For {\it Case 2} 
\begin{equation}
\tau - 1 = \frac{1}{\frac{\lambda^{2 \sigma_X}}{2 \vert A_2 \vert} -
  1} \, ,
\end{equation}
so $\tau \rightarrow \infty$ as $\lambda \rightarrow 2 \vert A_2 \vert$ and
since $A_2$ is different for each vortex the limit depends on which
vortex we are looking at.  Even though the quantitative features of
the scalings are different between the $(\zeta,\tau)$- and
$(X,Y)$-planes, qualitatively the fate of vortex pairs is the same in
that the distance between the members of each pair diverges as $\lambda \rightarrow 0$.  The
increase in distance and smearing of a single vortex pair can be seen
in Figure \ref{fig:StretchedVorts} by comparing image (a) to image
(b).  The ratio between the $\zeta$ and $\tau$ axes for each image is
kept constant, so the smearing of the region around the vortices is
not affected by the change in scale.

Bringing back $N$, we saw above that the scaling factor $\lambda^{\alpha}$ is replaced by $\lambda^{\alpha} N$. The question then arises, at what value of this scaling factor is the separation between the vortices and the antivortices  large enough so that they are visible? If we assume that in an experiment there is a value of the scaling total factor $\lambda^\alpha N = C$ below which they become distinguishable, then for a particular number of particles $N$ the parameter $\lambda$ must be tuned to values smaller than $(C/N)^{1/\alpha}$ for the individual vortices and antivortices to become visible.

\subsection{Effect of kick strength}

Here, we briefly show how the criticality of the cusp can be explored
without approaching the critical point of $V(x)$ by changing the
strength of the kick being applied.  If the kick has strength $Q$,
then $\delta(t) \rightarrow Q \delta(t)$ in our calculations.  The
result of this is that $p$ and $x$ are no longer treated on the same
footing because applying a stronger kick increases the ``momentum'' of
the system which causes the cusp to appear at earlier times.
Therefore, if we seek classically invariant coordinates where varying $Q$ or $\lambda$ only changes the quantum properties of the cusp, like Eq.\ (\ref{eq:wfscaled}), we must modify our previously defined classically invariant coordinates ($\zeta,\tau$). Suitable new coordinates are  $\tau = Q \lambda t$, $\zeta_x =
x/\sqrt{\lambda}$ and $\zeta_p = Q p/\sqrt{\lambda}$.  These
 result in the transformation $\lambda^\alpha \to
\lambda^\alpha Q^{2 \alpha - 3}$, so for {\it Case 1} and {\it Case 2} we have
$\lambda Q^{-1}$ and $\lambda^2 Q$, respectively, and we can achieve the same critical behaviour by varying $Q$ while fixing
$\lambda$.  The inverse relation of $Q$ between the two
cases arises because when the kinetic term is kicked ({\it Case 1})
with greater strength only amplitudes with small initial $p$ contribute to
the cusp until in the limit $Q \rightarrow \infty$ only $p_0 = 0$
contributes and the cusp vanishes.  The inverse limit for the kicked
potential term ({\it Case 2}) accomplishes the same thing because as $Q
\rightarrow 0$ the non-linearity, which is responsible for the cusp,
is removed.  Thus, systems with no phase transition at all can show the same
critical behaviour as a system with a second order phase transition by applying
weaker kicks.

\section{Non-$\delta$-Kick Quenches}
\label{sec:limitations}

\begin{figure}
	\begin{center}
		\includegraphics[width=1\columnwidth]{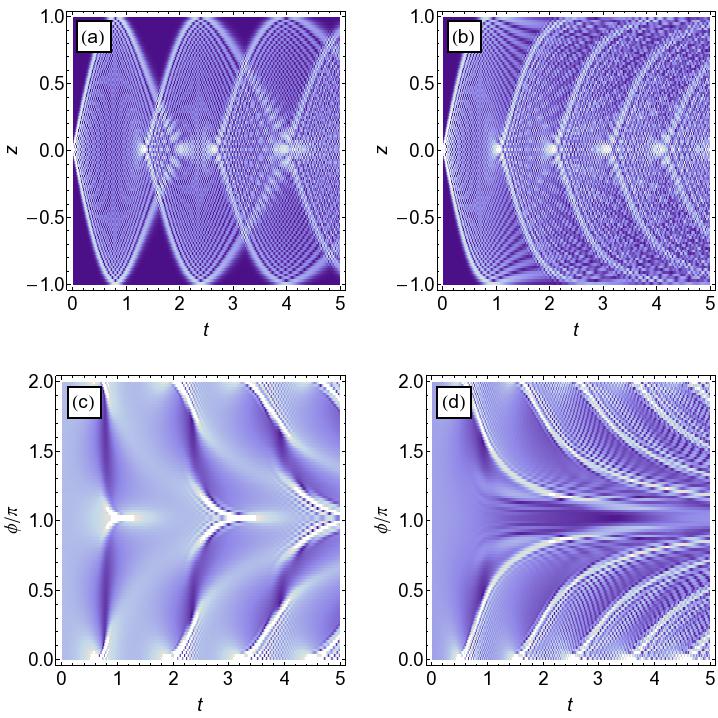}
	\end{center}
	\caption{The amplitude of the wave function for the two-mode Bose-Hubbard model with 100 bosons following a quench at $t=0$ where both terms in the Hamiltonian are present for the subsequent evolution. The initial state for all four panels is the Fock state with zero number difference, corresponding, for example, to a situation where two initially independent BECs are suddenly connected by a tunnelling barrier. The upper row gives the wave function in the number difference ($z$) basis and the lower row gives it in the phase difference ($\phi$) basis. Each column is for a different value of $\Lambda_{\mathrm{BH}}$:   the left column is for
          $\Lambda_{\mathrm{BH}} = 0.5$, so $\pi$-oscillations are
          possible and the right column is for
          $\Lambda_{\mathrm{BH}} = 1.5$ where $\pi$-oscillations are excluded.  The cusps created by the $\pi$-oscillations open toward the negative $t$ direction in both (a) and
      (c), the other cusps are due to plasma oscillations.  Note that Panel (a) is the quantum version
                  of Fig.\ \ref{fig:BJJ} when both Fig.\ \ref{fig:BJJ}(a) and Fig.\ \ref{fig:BJJ}(b) are laid on top of each other.} 
	\label{fig:pigrid}
\end{figure}

The $\delta$-kick quench allows for a simple analytic treatment and it also produces a single cusp, whereas for quenches where both terms in the Hamiltonian are present one typically gets oscillatory classical dynamics and hence multiple cusps, like in Figure \ref{fig:BJJ} and also in its quantum version Figure \ref{fig:pigrid}. The interference between the different cusps makes the quantum wave function more complicated, although one cusp will dominate in the immediate vicinity of its cusp point.   For these other types of quenches we do not expect the critical scaling to be the
same as the kicked cases, but we do still expect there to be some form of
scaling because this is a feature of the Pearcey function and the basic claim of catastrophe theory is that any structurally stable singularity must be mappable onto one of the canonical catastrophes.

In fact, we can still make some scaling arguments based on the results
from the kicked cases.  In deriving Eq.\ (\ref{eq:wfscaled}) we defined the new coordinates $\zeta =
x/\sqrt{\lambda} = p/\sqrt{\lambda}$ and $\tau = \lambda t$ which were
used to remove any classical scaling from the dynamics by making
Hamilton's equations scale invariant in $\lambda$.  The cusp
generating function, which represents the action, was not scale
invariant and the transformation resulted in $\Phi \rightarrow \lambda \Phi$
({\it Case 1}) and $\Phi \rightarrow \lambda^2 \Phi$ ({\it Case 2}).  One can proceed in a similar vein in the case of the full Hamiltonian $H=p^2/2+V(x)$, where the potential $V(x)=\lambda x^2 \pm x^4$, by looking for scalings of the classical coordinates that leave Hamilton's equations invariant. Hamilton's equations in this case are
\begin{eqnarray}
\dot{x}=p \\
\dot{p} = - 2 \lambda x \mp 4 x^3, 
\end{eqnarray}
and defining the new coordinates 
\begin{eqnarray}
\zeta_{x}=x/\sqrt{\lambda} \label{eq:nokickscale1} \\
\zeta_{p}=p/\lambda \\ 
\tau = \sqrt{\lambda} t \label{eq:nokickscale2}
\end{eqnarray}
 transforms them to
\begin{eqnarray}
\dot{\zeta}_{x}=\zeta_p \\
\dot{\zeta}_{p} = - 2 \zeta_{x}  \mp 4 \zeta_{x}^3, 
\end{eqnarray}
where the time derivative is now with respect to $\tau$. Plugging the new coordinates into the action $S=\int [p^2/2-V(x;\lambda)] \rmd t$ gives $S= \lambda^{3/2} \int [\zeta_{p}^{2}/2-V(\zeta_{x})]\rmd \tau$. Therefore, the action is transformed to $S \rightarrow \lambda^{3/2} S$.
Interestingly, the exponent, 3/2, is halfway between the exponents for
the two kicked cases signalling each term in the Hamiltonian is
playing an equal role in generating the dynamics.  Under these
transformations the propagator is
\begin{equation}
K(\zeta, \tau;\zeta_0, \tau_0) = \int \mathcal{D}[\zeta(\tau)] \rme^{\rmi
  N \lambda^{3/2} S[\zeta(\tau)]} \ .
\end{equation}
We shall not analyze the quantum dynamics this generates here, but we note that an analytic treatment of the wave function that is valid away from the immediate region of the cusp points has been given by one of us (DO) in reference \cite{odell12}. It uses a uniform approximation to extract the Airy function that decorates the fold lines that emanate from the cusp point.

Let us instead confine ourselves to a numerical solution obtained by an exact diagonalization of the full quantum Hamiltonian given in Eq.\ (\ref{eq:BH}) for the two-mode Bose-Hubbard model and consider its qualitative features.  The results are plotted in  Figure \ref{fig:pigrid} which shows the dynamics of the modulus of the wave function where the initial state is the single number difference (Fock) state $\vert 0 \rangle$, so at $t = 0$ the system has exactly $N/2$ bosons on each site.  The top row
shows the wave function in the number difference ($z$) basis where panel (a) is for
$\Lambda_{\mathrm{BH}} = 0.5$ and
represents the quantum version of the combined panels of Figure
\ref{fig:BJJ}.  Once again we can see the periodic cusps from plasma and
$\pi$-oscillations opening in the postive and negative $t$
directions, respectively.  Their combined interference pattern forms a periodic
diamond structure which grows in time.  Panel (b) shows the same
dynamics except $\Lambda_{\mathrm{BH}} = 1.5$, so the $\pi$-cusp vanishes.  Panels (c) and (d) show the wave function in the phase difference ($\phi$) basis for the same values of $\Lambda_{\mathrm{BH}}$ as in (a) and (b), respectively.
The periodic $\pi$-cusp is clearly visible at the centre 
of (c) bordered by half cusps from the plasma oscillations around
zero phase difference.  In (d) the plasma-cusps remain, but the $\pi$-cusps have vanished
due to the excited state phase transition.  
The $\phi$ basis is useful because  we can use the
potential in Eq.\ (\ref{eq:approxBH}) to give us the scaling of the
size and position of the cusps as $\lambda \rightarrow 0$, namely,
$t_{\mathrm{cusp}} \propto \lambda^{-1/2}$ and $\phi_{\mathrm{cusp}}
\propto \lambda^{1/2}$.  These scalings were already anticipated in Eqns.\ (\ref{eq:nokickscale1}) and (\ref{eq:nokickscale2}).

The main difficulty in numerically determining the scaling of the vortices' separation
comes from the interference with the plasma cusp, but the
scalings above can help to design a better initial state which shows
the cusps and their vortices more clearly.  One might consider using a
superposition of $\hat{\phi}$-states around $\pi$ instead of the
$\vert 0 \rangle$ $\hat{z}$-state which inconveniently gives a broad
superposition over all $\hat{\phi}$-states. Finally, we note that in the exact solution plotted in Figure \ref{fig:pigrid}, Fock space is discrete and this can smear out the vortex cores making
their positions difficult to track.  However, this discretization shrinks
with increasing $N$, becoming invisible for a large enough system.

\section{Discussion and Conclusion}
\label{sec:conclusion}

The main message of this paper is that close to a singularity the wave function takes on a universal form, namely one of the structures predicted by catastrophe theory. These catastrophes obey scaling laws and also occur generically during dynamics without the need for fine tuning. This means we expect them to occur in a wide variety of situations, as is the case in optics, through analogues of the phenomenon of natural focusing. Of course, in high symmetry situations catastrophes can reduce to simpler structures (e.g.\ points rather than cusps) but these unfold to one of the canonical catastrophes when that symmetry is broken. We therefore come to the perhaps counter-intuitive conclusion that singularities represent islands of predictability in a sea of complexity, acting as organizing centres around which the wave function can only take on one of a limited number of forms and has well defined properties.

In previous work \cite{odell12}, we showed that in many-particle problems wave catastrophes occur in Fock space. They  are naturally discretized by the granularity of the particles but become singular in the mean-field limit where the discretization is neglected. In this paper we worked within the continuum approximation where the granularity is neglected, but in contrast to the mean-field approximation the essential quantum nature of the number and phase operators is preserved as encapsulated by the commutation relation $[\hat{\phi},\hat{z}]=2 \mathrm{i} /N$. Furthermore, we specialized to the case of a $\delta$-kick quench as this allows us to analytically solve for the Fock-space wave function of two-mode problems and represent it as a Pearcey function which is the universal wave function associated with cusp catastrophes. In particular, the centrepiece of our analysis is the result given in Eq.\ (\ref{eq:wfscaled}) which shows how the wave function scales with a parameter $\lambda$ which controls a second-order dynamical phase transition: the scaling exponents for various properties of the wave function are summarized in Table \ref{tab:CE} and include both classical (mean-field) aspects such as the position and size of the cusp as well as quantum (many-particle) aspects such as the amplitude of the interference pattern and its fringe spacing  in different directions.

A physical example where this general two-mode wave function applies is to the two-mode Bose-Hubbard model where there is a dynamical phase transition describing the appearance/disappearance of $\pi$-oscillations. Since our treatment is based on a general quartic potential (where $\lambda$ controls the size of the quadratic term), it can be applied to other dynamical phase transitions too. The classical scaling of the cusp is independent of which term in the Hamiltonian is kicked, but when we go to the quantum theory kicking the potential term results in the cusp being more sensitive to changes in the control parameter $\lambda$ as compared to when the kinetic term is kicked. As the phase transition is approached ($\lambda \rightarrow 0$) the cusp appears at later times and also shrinks, i.e.\ grows more slowly with time.  The quantum aspects of the scaling mean that the interference peaks become fainter and farther apart as $\lambda \rightarrow 0$.  When we explicitly include the number of particles $N$ in the theory we find that the scaling parameter  is transformed to $\lambda \rightarrow \lambda N$ and there is therefore a clash of limits between the phase transition as $\lambda \rightarrow 0$ and the thermodynamic limit $N \rightarrow \infty$. 

Apart from its scaling properties, another important feature of the Pearcey function is a network of vortex-antivortex pairs inside the cusp. When applied to many-particle dynamics this implies that there are vortex-antivortex pairs in the two-dimensional plane given by Fock space plus time. As far as we are aware the observation that there can be topological structures in such spaces, which are the Hilbert spaces describing many-particle quantum systems, is new and warrants further investigation. In the present context we find that as the phase transition is approached the vortex-antivortex pairs are pulled apart in an anisotropic manner described by the two Berry indices.

A key question is whether the present analysis can be applied to more complicated many-particle systems. In the three mode case (corresponding, e.g.,  to the three-site Bose Hubbard model) the control space is three dimensional (two dimensional Fock space plus time) and following a quench one indeed finds $K=3$ catastrophes (swallowtail, elliptic umbilic, hyperbolic umbilic) \cite{yee}. In principle one can continue on to more modes and hence to higher catastrophes but the increasing complexity of the catastrophes as $K$ becomes large would make this a challenging task for even a moderately sized lattice of sites as there is essentially too much information. A more promising approach in this case would be to switch to a statistical version of catastrophe theory where the statistics of the fluctuations of the wave function are the central objects of interest \cite{berry77}. 
 
\ack

We thank Maxim Olshanii for discussions and
 the Natural Sciences and Engineering Research Council
of Canada for funding. 

\appendix

\section{Derivation of scaled wave function}
\label{app:derivation}

Here, we explicitly go through the steps in deriving Eq.\
(\ref{eq:wfscaled}) for {\it Case 2} (kicked potential) starting with Eq.\ (\ref{eq:wf2}) (the derivation for  {\it Case 1} is
similar).  To simplify the notation
we will ignore all numerical factors and overall phases.  We start by substituting Eq.\
(\ref{eq:approxpot}) with $a_0 = 0$, $a_2 = \lambda/2$, $a_3 = 0$ and
$a_4 = \pm 1$ into Eq.\ (\ref{eq:wf2})
\begin{equation}
\psi_2 (x, t; \lambda) \propto \frac{1}{\sqrt{t}}
\int_{-\infty}^{\infty} \rmd x_0 \rme^{\rmi \left [ -\frac{x x_0}{t}
    +(1 - \lambda t) \frac{x_0^2}{2 t} \mp x_0^4  \right ]} \, . 
\end{equation}
We then substitute in the rescaled position and time variables, $\tau =
\lambda t$ and $\zeta = x/\sqrt{\lambda}$ so the cusp is stationary
with respect to $\lambda$ in the rescaled plane giving
\begin{eqnarray}
\psi_2 (\zeta, \tau; \lambda) &\propto& \sqrt{\frac{\lambda}{\tau}} 
\int_{-\infty}^{\infty} \rmd x_0 \rme^{\mp \rmi \left [ \pm
    \lambda^{3/2} \frac{\zeta x_0}{\tau} \mp \lambda (1 - \tau) \frac{x_0^2}{2 \tau}
    + x_0^4  \right ]} \nonumber \\
&=& \sqrt{\frac{\lambda}{\tau}} 
\int_{-\infty}^{\infty} \rmd x_0 \rme^{\mp \rmi \left [ \pm
    \lambda^{3/2} Y_2(\zeta, \tau) x_0 \mp \lambda X_2(\tau) x_0^2
    + x_0^4  \right ]} \nonumber \\
&=& \sqrt{\frac{\lambda}{\tau}} \, \mathrm{Pe} \left [ \mp \lambda
  X_2(\tau), \lambda^{3/2} Y_2(\zeta, \tau) \right ]
\label{eq:pe}
\end{eqnarray}
where we have used the fact that $\mathrm{Pe} \left [ X, -Y \right ] =
  \mathrm{Pe} \left [ X, Y \right ]$.  We can see Eq.\ (\ref{eq:pe})
  matches Eq.\ (\ref{eq:wfscaled}) for the kicked potential case
  ($\alpha = 2$) given the Arnold index, $\beta = 1/4$, and Berry
  indices, $\sigma_X = 1/2$ and $\sigma_Y = 3/4$.  The $\mp$ sign
  indicates whether the quartic term in the potential is positive or
  negative, respectively.

\end{document}